\documentclass[aps, prd
, twocolumn
, nofootinbib
, superscriptaddress
, preprintnumbers
]{revtex4-2}

\usepackage{graphicx}
\usepackage{xcolor}
\usepackage{mathrsfs,mathtools}
\usepackage{physics,amssymb}
\usepackage{siunitx}
\usepackage{bm}
\usepackage{soul}
\usepackage{cancel}
\usepackage{url}
\usepackage{longtable}
\usepackage{xspace}
\usepackage{acronym, comment}
\usepackage{aas_macros}
\usepackage[colorlinks=true
,urlcolor=DARKBLUE
,anchorcolor=DARKBLUE
,citecolor=DARKBLUE
,filecolor=DARKBLUE
,linkcolor=DARKBLUE
,menucolor=DARKBLUE
,linktocpage=true
,pdfproducer=medialab
,pdfa=true
]{hyperref}

\DeclareMathOperator{\arcsinh}{asinh}
\DeclareMathOperator{\arccosh}{acosh}

\newcommand{\ee}{\mathrm{e}}
\newcommand{\Mpl}{M_\mathrm{Pl}}

\newcommand{\uth}{\mathrm{th}}
\newcommand{\amp}{\mathrm{amp}}
\newcommand{\peak}{\mathrm{peak}}
\newcommand{\GW}{\mathrm{GW}}

\newcommand{\ie}{i.e.}
\newcommand{\eg}{e.g.}

\newcommand{\ui}{\mathrm{i}}

\newcommand{\um}{\mathrm{m}}
\newcommand{\ut}{\mathrm{t}}

\newcommand{\calO}{\mathcal{O}}

\newcommand{\calP}{\mathcal{P}}

\newcommand{\uU}{\mathrm{U}}
\newcommand{\calV}{\mathcal{V}}
\newcommand{\bfx}{\mathbf{x}}

\newcommand{\beae}[1]{\begin{equation}\begin{aligned} #1 \end{aligned}\end{equation}}
\newcommand{\bege}[1]{\begin{equation}\begin{gathered} #1 \end{gathered}\end{equation}}
\newcommand{\bae}[1]{\begin{align} #1 \end{align}}

\newcommand{\bfe}[4]{
\begin{figure} 
	\centering
	\includegraphics[#1]{#2}
	\caption{#3}
	\label{#4}
\end{figure}}
\newcommand{\bme}[1]{\begin{multline} #1 \end{multline}}
\newcommand{\bmte}[1]{\begin{multlined}[t] #1 \end{multlined}}

\definecolor{MONZA}{HTML}{CF000F}
\definecolor{DARKBLUE}{HTML}{00008b}
\definecolor{DARKMAGENTA}{HTML}{8b008b}
\definecolor{PURPLE}{HTML}{8822FF}

\begin{document}
\title{Quintessence with tachyonic resonance and \\ late-time cosmic-microwave-background and gravitational-wave signals}
\date{\today}

\author{Shun Yoshioka}
\email{yoshioka.shun.k6@s.mail.nagoya-u.ac.jp}
\affiliation{Department of Physics, Nagoya University, 
Furo-cho Chikusa-ku,
Nagoya 464-8602, Japan}

\author{Kiyotomo Ichiki}
\email{ichiki.kiyotomo.a9@f.mail.nagoya-u.ac.jp}
\affiliation{Department of Physics, Nagoya University, 
Furo-cho Chikusa-ku,
Nagoya 464-8602, Japan}
\affiliation{Kobayashi-Maskawa Institute for the Origin of Particles and the Universe, 
Nagoya University, 
Nagoya 464-8602, Japan}
\affiliation{Institute for Advanced Research, Nagoya University,
Furo-cho Chikusa-ku, 
Nagoya 464-8601, Japan}

\author{Yuichiro Tada}
\email{yuichiro.tada@rikkyo.ac.jp}
\affiliation{Department of Physics, Rikkyo University, Toshima, Tokyo 171-8501, Japan}
\affiliation{Institute for Advanced Research, Nagoya University,
Furo-cho Chikusa-ku, 
Nagoya 464-8601, Japan}
\affiliation{Department of Physics, Nagoya University, 
Furo-cho Chikusa-ku,
Nagoya 464-8602, Japan}

\author{Takahiro Terada}
\email{terada@eken.phys.nagoya-u.ac.jp}
\affiliation{Kobayashi-Maskawa Institute for the Origin of Particles and the Universe, 
Nagoya University, 
Nagoya 464-8602, Japan}

\begin{abstract}
Combinations of recent cosmological observations, including \ac{DESI}, show hints of a dynamical nature for dark energy.  While the data suggest the possibility of the phantom crossing, it is worth thoroughly exploring quintessence models.  Given that phenomenological parametrisations of the equation-of-state parameter $w(a)$ with a sharp transitional feature fit the data well, we study the realisation of such models in quintessence.  In the late Universe, the quintessence field begins to oscillate abruptly, changing the behaviour of $w$.  Naturally, such a model entails tachyonic instability, and particle production modifies $w$.  We perform numerical lattice simulations to study the time dependence of $w$.  In addition, the violent particle production produces significant density perturbations and the stochastic gravitational-wave background, whose characteristic scale depends on the mass scale of the quintessence around the minimum of the potential.  
We discuss the observability of these late-time cosmological signals through cosmic microwave background, quasar astrometry, pulsar timing arrays, and other observational probes. 
\end{abstract}

\preprint{RUP-26-2}
\maketitle

\acrodef{EoS}{equation-of-state}
\acrodef{EoM}{equation of motion}
\acrodef{DESI}{Dark Energy Spectroscopic Instrument}
\acrodef{BAO}{baryon acoustic oscillation}
\acrodef{CMB}{cosmic microwave background}
\acrodef{MCMC}{Markov chain Monte Carlo}
\acrodef{ACT}{Atacama Cosmology Telescope}
\acrodef{SN}{supernova}
\newacroplural{SN}[SNe]{supernovae}
\acrodef{CPL}{Chevallier--Polarski--Linder}
\acrodef{DSCh}{Dutta--Scherrer--Chiba}
\acrodef{TDE}{transitional dark energy}
\acrodef{GW}{gravitational wave}
\acrodef{ISW}{integrated Sachs--Wolfe}
\acrodef{LISW}{late-time ISW}
\acrodef{PTA}{pulsar timing array}
\acrodef{IPTA}{International Pulsar Timing Array}
\acrodef{SKA}{Square Kilometre Array}

\acresetall
\section{Introduction}\label{sec:intro}

Dark energy, the origin of the recent accelerated expansion of the Universe, is viewed as one of the most profound mysteries of the Universe. 
Recently, \ac{DESI} reported evidence~\cite{DESI:2024mwx, DESI:2024hhd, DESI:2025zgx, DESI:2025fii} of the time dependence of dark energy by combining their data with those of \ac{CMB}~\cite{Planck:2018nkj, Planck:2019nip, ACT:2023kun, ACT:2023oei} and of \acp{SN}~\cite{Brout:2022vxf, DES:2024tys, Rubin:2023ovl}. Their \ac{BAO}~\cite{DESI:2024mwx} and full-shape~\cite{DESI:2024hhd} analyses at data release 1 (DR1) as well as the \ac{BAO} analysis at DR2~\cite{DESI:2025zgx, DESI:2025fii} are mutually consistent.  Interestingly, the fact that the $\Lambda$CDM (cold dark matter; CDM) model is relatively disfavored does not significantly depend on the specific choice of the parametrization of the \ac{EoS} parameter $w(a)$ such as the \ac{CPL} parametrization~\cite{Chevallier:2000qy, Linder:2002et}, which is a two-parameter extension of the $\Lambda$CDM model, in the sense that other smooth two-parameter phenomenological extensions lead to qualitatively similar results~\cite{DESI:2025fii}.  Moreover, the fit does not significantly improve beyond the two-parameter extensions~\cite{DESI:2025fii}, so they may carry some essential features of the data.  These phenomenological fits by the extended models typically favour a phantom regime, \ie, with $w < - 1$ at a high redshift range. 
However, the uncertainties on the \ac{EoS} parameter are relatively large at such a redshift range, so the phantom crossing of the dark energy has not been established. 
For a recent status of dynamical dark energy, see a review~\cite{Capozziello:2025qmh}.

Time dependence of dark energy and general covariance suggest that the dynamical dark energy is represented by a scalar field $\phi$, called quintessence in this context.  For a canonical scalar field, they satisfy the null energy condition, $w \geq -1$, while a phantom regime $w < -1$~\cite{Caldwell:1999ew} would require either a negative-norm ghost or an instability~\cite{Carroll:2003st, Cline:2003gs, Vikman:2004dc}.  In this paper, we consider only non-phantom (\ie, quintessence) fields.\footnote{
It is possible to effectively mimic the phantom behaviour in the presence of interactions between dark energy and dark matter~\cite{Huey:2004qv, Das:2005yj, Chakraborty:2025syu, Khoury:2025txd}.  Realisation of phantom degrees of freedom for other setups was also discussed in the literature. Dozens of examples are cited in Refs.~\cite{DESI:2025fii}. See also more recent Refs.~\cite{Tsujikawa:2025wca, Chen:2025ywv, Tsujikawa:2026xqm}. 
} See Refs.~\cite{Tada:2024znt, Yin:2024hba, Berghaus:2024kra, Ramadan:2024kmn, Notari:2024rti, Gialamas:2024lyw, Ye:2024ywg, Wolf:2024eph, Wolf:2024stt, Bhattacharya:2024kxp, Notari:2024rti, Berbig:2024aee, Aboubrahim:2024cyk, Shajib:2025tpd, Wolf:2025jlc, Borghetto:2025jrk, Luu:2025fgw, Nakagawa:2025ejs, Urena-Lopez:2025rad, Wolf:2025jed, Murai:2025msx, Lin:2025gne, Gialamas:2025pwv, Wolf:2025acj, Bouhmadi-Lopez:2026wub, Ibitoye:2026whe} on the quintessential interpretation of the dynamical dark energy. 
In analogy to inflationary physics, typical quintessence models have smooth scalar potentials leading to a smooth curve $w(a)$.  Scalar potentials of such models can be Taylor-expanded around a relevant point in the field space, and the resultant cosmology can be studied by using the formalism of Dutta and Scherrer~\cite{Dutta:2008qn} and its generalisation by Chiba~\cite{Chiba:2009sj} (\acsu{DSCh}).  

However, not all the quintessence models fall in this category. As an alternative possibility, dark energy with sharp transitional feature(s) like a kink~\cite{Notari:2024rti} and a step~\cite{Bassett:2002qu, Corasaniti:2002vg, Gialamas:2024lyw, Keeley:2019esp, Keeley:2025stf, Pang:2024qyh, Payeur:2024dnq} (see also Ref.~\cite{Cheng:2025yue}) in the \ac{EoS} parameter as a function of time (scale factor or redshift) was also studied.  In particular, a dark energy model with a step-like feature is known as \ac{TDE}~\cite{Keeley:2019esp, Keeley:2025stf}. As we will see in Sec.~\ref{sec:toy_model}, the \ac{TDE} model can fit the observational data as well as the \ac{DSCh} model, which is a proxy of typical, simple, and smooth quintessence models. Then, it is legitimate to ask what the microphysics origin of the \ac{TDE} model is. 

In this paper, we discuss the approximate realisation of \ac{TDE}~\cite{Keeley:2019esp, Keeley:2025stf} in terms of quintessence.  It turns out that such quintessence models have an essential feature involving particle production and generation of significant inhomogeneities and strong \acp{GW} as we will discuss in detail.  This challenges the following two conventional views: (1) significant modification of cosmology at a late time is hard because we can constrain well the part of the universe close to us, and (2) characteristic physical scales in dark energy models are comparable with the present Hubble scale.  In fact, we will discuss $\mathcal{O}(1)$ transmutation of dark energy density into the energy density of a dark-matter-like component, generation of $\mathcal{O}(1)$ perturbations, and generation of the stochastic \ac{GW} background from such enhanced perturbations, all of which occurs at a very late time $z \approx \mathcal{O}(0.1)$.  The typical wavenumber scale, corresponding to the typical frequency of the \acp{GW}, can be much higher than the Hubble scale and well below the ultraviolet (UV) cutoff scale of the model. Interestingly, all these novel cosmological late-time features originate from the attempt to explain the dynamical dark energy. 

The organisation of the paper is as follows.  In Sec.~\ref{sec:toy_model}, we introduce the \ac{TDE} as a phenomenological toy model and discuss its observational status. The comparison of the \ac{TDE} model with the DSCh model is given in Appendix~\ref{sec: FUll MCMC}. As we will see, the two types of models have comparable ability to fit the data. While the typical or sufficiently smooth quintessence model is well studied, the physical basis of the \ac{TDE} model has not been explored much. Therefore, we study a quintessential realisation of the \ac{TDE} model in Sec.~\ref{sec:quintessence}, where we find rich observational consequences. While we present the results of numerical lattice simulations in the section, we complement it with the linear analyses of perturbations in Appendix~\ref{sec: linear analysis}. Some analytic estimates regarding the initial stage of the transition are given in Appendix~\ref{sec: estimate_period}.   Our conclusions are given in Sec.~\ref{sec:conclusion}. 
We adopt the reduced-Planck unit where $c$, $\hbar$, $8\pi G=1/\Mpl^2$, and $k_\mathrm{B}$ are set to unity unless explicitly denoted.

\section{Transitional Dark Energy}\label{sec:toy_model}

Let us first introduce phenomenological dark energy models with transitional features. 
We consider a step model~\cite{Gialamas:2024lyw} defined by the following ansatz for the \ac{EoS} parameter\footnote{When the present time is sufficiently close to the transition in the step model, the finite transition time should not be neglected. In such a case, it will be more natural to consider a broken-linear model 
\begin{align}
    w(a) =
    \begin{cases}
    -1 & (a < a_\text{t}), \\
    -1 +  (w_0 + 1) \frac{a - a_\text{t}}{1 - a_\text{t}} & (a \geq a_\text{t}),
    \end{cases} \label{broken-linear_model}
\end{align}
where again $w_0 = w(1)$. In this case, $w(a)$ is continuous, but its derivative is not at $a = a_\text{t}$.  
A similar model in terms of redshift $z$ was studied in Ref.~\cite{Notari:2024rti}, where its realisation in a quintessence model was also given, albeit with a singularity in the scalar potential.
Our main interest is the step(-like) model since it is more generically realised in quintessence models, as we will see in the next section, though the broken-linear model can also be approximately realised in some parameter space without introducing a singularity in the potential.  
}
\begin{align}
    w(a) = \begin{cases}
        w_\infty  & (a < a_\text{t}), \\
        w_0 & (a \geq a_\text{t}),
    \end{cases} 
    \label{eq: TDE_EoS}
\end{align}
where $w_0$ and $w_\infty$ are the late-time and early-time values of $w$, respectively, 
and $a_\text{t}$ is the value of the scale factor at which the transition occurs from $w = w_\infty$ to $w = w_0$. Since we do not consider the phantom regime, we set $w_\infty = -1$ with the thawing quintessence in mind, while we keep $w_0$ as a free parameter.  
This step model can be viewed as a special case of the \ac{TDE} model~\cite{Bassett:2002qu, Corasaniti:2002vg, Keeley:2019esp, Keeley:2025stf}, where the step function is smoothened by the 
hyperbolic tangent function. In this paper, we use the terminology in which the \ac{TDE} includes the step model~\eqref{eq: TDE_EoS} and its approximations, not limited to those with 
the hyperbolic tangent.  We show in the next section that the transition can be approximately realised in a quintessence model.

For comparison, let us also introduce a setup without the sudden transitional feature. Generic quintessence models have a smooth scalar potential, and slow-roll dynamics are usually assumed to mimic the cosmological constant.  Then, it is often the case that it suffices to consider the Taylor expansion of the potential.  The advantage of this approach is that the details of the scalar potential are not important, and a few derivatives of the potential are enough to approximately solve the dynamics of the quintessence. 
Analytic approximate formulas were obtained using the information up to and including the second derivative of the potential~\cite{Dutta:2008qn, Chiba:2009sj}, dubbed the \ac{DSCh} formalism.  
The formalism is characterised by three parameters: the present \ac{EoS} parameter $w_0$, the density parameter $\Omega_\phi$ ($=1-\Omega_\um$ with the matter density $\Omega_\um$, supposing the flat universe), and a parameter $K$ that controls the time evolution of the scale factor. The time-evolution of the \ac{EoS} parameter is given by
\begin{align}
    w(a) = - 1 + (1 + w_0) a^{3(K-1)} \mathcal{F}(a\mid K,\Omega_\phi), \label{DSCh}
\end{align}
where the function of the scale factor, $\mathcal{F}(a\mid K,\Omega_\phi)$, depends on $K$ and $\Omega_\phi$; see 
Refs.~\cite{Dutta:2008qn, Chiba:2009sj} for its explicit expression. 
The parameter $K$ is related to the derivatives of the scalar potential as $K \equiv \sqrt{1 - 4 V''(\phi_\text{i}) / 3 V(\phi_\text{i})}$, where $\phi_\text{i}$ is the initial value of the quintessence field $\phi$, $V(\phi)$ is its potential, and a prime denotes the derivative with respect to its argument.

\subsection{MCMC analysis}
\label{sec: MCMC analysis}

To study how competitive the \ac{TDE} model is compared to the \ac{DSCh} model, we perform a \ac{MCMC} analysis. 
We modify the cosmological code \texttt{CAMB}~\cite{Lewis:1999bs,Howlett:2012mh}, to implement the behaviour of dark energy models. 
For each model, we perform a \ac{MCMC} analysis using \texttt{cobaya}~\cite{Torrado:2020dgo, Lewis:2002ah, Lewis:2013hha} including the data from the \ac{DESI} DR2 \ac{BAO} measurements~\cite{DESI:2025zgx}, the \emph{Planck} \ac{CMB} measurements~\cite{Aghanim:2018oex, Aghanim:2019ame}, and the \acp{SN} measurements from PantheonPlus~\cite{Brout:2022vxf}, DES-SN5YR~\cite{DES:2024tys},\footnote{See Ref.~\cite{DES:2025sig} for a reanalysis, which leads to a weaker evidence of dynamical dark energy.} and Union3~\cite{Rubin:2023ovl}. 
For \ac{CMB} data, we use Planck 2018 low-$\ell$ temperature and polarisation data, Planck 2018 high-$\ell$ TTTEEE data, and Planck 2018 lensing data.
Posterior distributions are generated using \texttt{GetDist}~\cite{Lewis:2019xzd}.

The MCMC analysis is conducted using six fiducial cosmological parameters, 
$\Omega_\mathrm{b}h^2$, $\Omega_\mathrm{c}h^2$, $100\theta_\mathrm{MC}$, $\log(10^{10} A_\mathrm{s})$, $n_\mathrm{s}$, and $\tau_\mathrm{reio}$, along with two additional parameters for each case ($w_0$ and $a_\text{t}$ for the TDE model and $w_0$ and $K$ for the DSCh model). Flat priors are assumed for all parameters, with the ranges $w_0 \in (-2,1)$ and $a_\mathrm{t} \in (0.2,1)$ for the TDE model (see Appendix~\ref{sec: FUll MCMC} for the prior of the DSCh model). 

\begin{table}
\renewcommand{\arraystretch}{1.3}
\footnotesize
\begin{center}
  \caption{$68\%$ CL posterior for TDE model. Best-fit parameters are listed in small text. $H_0$ is normalised by $(\si{km/s/Mpc})$. }
  \label{tab:MCMC_TDE}
  \begin{ruledtabular}
  \begin{tabular}{
    lcccc}
     & & DESY5 & PantheonPlus & Union3 \\
    \hline
    $w_0$ & & $ -0.32^{+0.18}_{-0.61}$ & $ -0.54^{+0.21}_{-0.51}$& $ -0.548^{+0.076}_{-0.40}$ \\
     & & {\scriptsize$-0.31602593$} &{\scriptsize$ -0.92981337$} & {\scriptsize$ -0.65243413$} \\
    $a_\ut$ & & $ 0.913^{+0.062}_{-0.0054}$ & $ > 0.853$ & $ 0.861^{+0.10}_{-0.032} $ \\
     & & {\scriptsize$0.95172941$} & {\scriptsize$ 0.87095378$} & {\scriptsize$ 0.85634389$} \\
    $\Omega_\mathrm{b} h^2$ & & $ 0.02255\pm 0.00013$ & $ 0.02255\pm 0.00013$ & $ 0.02256\pm 0.00013$ \\
     & & {\scriptsize$0.022621533$} & {\scriptsize$ 0.022563837$} & {\scriptsize$ 0.022549837$} \\
    $\Omega_\mathrm{c} h^2$ & & $ 0.11760\pm 0.00066$ & $ 0.11766\pm 0.00068$ & $ 0.11752\pm 0.00068$ \\
     & & {\scriptsize$0.11733714$} &{\scriptsize$ 0.11795790$} & {\scriptsize$ 0.11789140$} \\
    \hline
    $H_0$ & & $ 65.97^{+0.88}_{-0.71}$ & $ 67.36\pm 0.74$& $ 65.8\pm 1.2$ \\
     & & {\scriptsize$66.332425$} & {\scriptsize$ 67.705310$} & {\scriptsize$ 65.171810$} \\
    $ \Delta \chi^2$ & & $ -14.8$& $ -3.66$ & $ -8.68$ \\
  \end{tabular}
\end{ruledtabular}
\end{center}
\end{table}
The best-fit parameters and marginalized $68 \%$ posterior distributions are summarised in Table~\ref{tab:MCMC_TDE}.
The full posterior distributions are shown in Appendix~\ref{sec: FUll MCMC}. 
The current value of the Hubble parameter, $H_0$, is derived from the other parameters. The difference of the minimum chi-squared value from that of the $\Lambda$CDM model, $\Delta \chi^2$, is also shown to indicate the preference of the model.

These preferences 
significantly depend on the SN data set.
It has been discussed that a magnitude offset $\sim\SI{0.04}{mag}$ between the low and high redshift \ac{SN} data in the DES5Y compilation, pointed out in Ref.~\cite{Efstathiou:2024xcq}, may be a possible source of such low-redshift systematics. Ref.~\cite{Notari:2024zmi} shows that marginalising such an offset in DES5Y weakens the evidence for dynamical dark energy (see also Refs.~\cite{RoyChoudhury:2024wri,Huang:2025som}).
The DES collaboration had, however, argued against these treatments of the offset~\cite{DES:2025tir}.
For other diagnostics of and conservative views on dynamical dark energy, see, e.g., Refs.~\cite{Luongo:2024fww, Carloni:2024zpl, Dinda:2024kjf}.

As seen in Appendix~\ref{sec: FUll MCMC}, the \ac{TDE} and \ac{DSCh} models exhibit similar levels of preference, and 
the posterior distributions of $w(a)$ are similar across different models.
It is hence interesting to ask about the underlying physical models.  For the DSCh case, by construction, it is an approximation to a generic class of quintessence models with a smooth scalar potential.  On the other hand, the physical status of the TDE model is apparently less clear. The purpose of this paper is to present a physical realization to such a model with the transitional feature and discuss its observational implications. 

Note that, in the TDE model, the best-fit transition point of the scale factor $a_\mathrm{t}\sim 0.9$ (or the redshift $z_\mathrm{t} \sim 0.1$) differs from that of the CPL model, for which the phantom-crossing redshift is typically $z_\mathrm{t} \sim 0.5$~\cite{DESI:2025fii}. 
This shift arises because the CPL model allows for the phantom behaviour, whereas the TDE model avoids this region, resulting in different values of $a_\text{t}$. 
This suggests that the preference for the \ac{TDE} arises mostly from the \ac{SN} data.

In the rest of the paper, we focus on the TDE model with fixed parameters: $w_0 \sim -0.5$ and $a_\mathrm{t} \sim 0.9$.

\section{Resonance and observational signatures of a realistic transitional dark energy model}\label{sec:quintessence}

\subsection{Sudden-roll quintessence as transitional dark energy and its resonant feature}

The TDE discussed in the previous section can be approximately realised by quintessence. Suppose that, initially, the quintessence field slowly rolls down its flat potential. This phase corresponds to the initial quasi-de-Sitter phase with $w_\infty \simeq -1$. Suppose, additionally, that the rolling of the scalar field is rapidly accelerated by a smooth but local feature in its potential, followed by a non-slow-roll phase with the \ac{EoS} parameter $w_0$.  If the intermediate phase is short enough, the \ac{EoS} parameter changes abruptly, thereby approximately reproducing the step model. 
If the rolling of the scalar field were instantaneous, it would correspond to the step model in Eq.~\eqref{eq: TDE_EoS}. We do not consider such a limit and consider a rapid but finite-time transition as specified below.

To be concrete, let us consider a scalar quintessence field $\phi$ with a plateau potential, whose Lagrangian density is given by\footnote{The model is known as the T-model $\alpha$-attractor in the context of primordial inflation~\cite{Kallosh:2013hoa, Kallosh:2013yoa, Carrasco:2015pla}. The asymptotic shift symmetry of $\phi$ at $|\phi| \gg f$ can be interpreted, e.g., as a manifestation of the scale invariance as in Higgs inflation~\cite{Bezrukov:2007ep} and $R^2$ inflation~\cite{Starobinsky:1980te}, or alternatively a $\uU(1)$ symmetry in the pure natural inflation~\cite{Nomura:2017ehb}. }
\beae{
    &\mathcal{L}= - \frac{1}{2}g^{\mu\nu}\partial_\mu \phi \partial_\nu \phi - V(\phi), \\
    &V(\phi)=V_0 \tanh^{2n} \left( \frac{\phi}{f} \right), \quad (n=1,2,3,\ldots), \label{quintessence_model}
}
while we focus on the case $n=1$ in the following analysis.
Here, $V_0$ is the overall factor, and the decay constant $f$ characterises the mass of the scalar at the origin as $V''(0)=2V_0/f^2\eqqcolon2m_\uth^2$, where the prime denotes the derivative with respect to the argument.\footnote{The factor-two in the definition of the typical mass scale $m_\uth$ is a convention~\cite{Tomberg:2021bll}.}  

Consider its background dynamics, where the scalar depends only on time as $\phi(t,\bfx)=\phi_0(t)$. 
If it is too deep on the plateau, $\abs{\phi_0}\gg f$, it is frozen because of the Hubble friction, and it behaves almost as an effective cosmological constant reproducing the $\Lambda$CDM model, which is not suitable for our purpose. 
If it is in the vicinity of the edge of the plateau $\abs{\phi_0}\gtrsim f$, it initially slow-rolls, and the \ac{EoS} parameter $w$ is almost $-1$ in the early universe.  As the Hubble parameter decreases well and $|\phi_0|$ approaches the edge of the plateau, however, it rolls down to and oscillates around the bottom of the potential. 
These dynamics are rapid (faster than the Hubble rate) for $f \ll \Mpl$. 
Then, the rapid oscillations of $\phi_0$ also lead to the rapid oscillations of $w$ (see Fig.~\ref{fig: lattice_EoS}).  
While these oscillate rapidly compared to the Hubble time scale, the change of the EoS parameter is observationally probed via the Hubble expansion, so it is useful to consider a time average over an oscillation, leading to an effectively constant \ac{EoS} parameter $\sim w_0$.\footnote{\label{footnote: w in n}
At the homogeneous level and with the average over the field oscillations, one can compute the \ac{EoS} parameter as a function of the dark energy density $\rho_\phi$ and hence $\phi$ using the technique of Refs.~\cite{Karam:2021sno, Tomberg:2021bll}. For $n=1$, an explicit formula is given in Appendix~A of Ref.~\cite{Matsui:2023ezh}. After the amplitude of $\phi$ decreases sufficiently so that the anharmonic effect becomes negligible, the time-averaged \ac{EoS} parameter reads $\expval{w}=(n-1)/(n+1)$ (see, e.g., Ref.~\cite{Cembranos:2015oya} and references therein).
However, as the matter component is still non-negligible today, and the perturbation resonance backreacts on the background dynamics, the current \ac{EoS} parameter $w_0$ is not necessarily given by this value in our setup. We will read off $w_0$ from the lattice simulation results.} 
In this way, $\phi$ traverses a flat part, concave (tachyonic) regions, and the convex region, so the potential cannot be approximated by a single quadratic function.  This means that the DSCh approximation does not apply to our potential. 

This is not the end of the story.  
In the context of inflation, this model with a small decay constant $f\ll \Mpl$ is known~\cite{Lozanov:2017hjm, Tomberg:2021bll} to exhibit a preheating phase during which the inflaton repeatedly climbs the potential up to the region where $V''<0$, where tachyonic instability occurs. 
The repetition of the tachyonic instability leads to structures like resonance bands, and it is called tachyonic resonance~\cite{Dufaux:2006ee, Abolhasani:2009nb} (also called flapping resonance in Refs.~\cite{Kitajima:2018zco, Fukunaga:2019unq}). 
That is, certain initial fluctuations $\delta\phi$ on top of the homogenous background $\phi_0$, even due to the quantum vacuum fluctuations, can exponentially grow right after the onset of the oscillations and possibly change the dynamics on average.

Qualitatively similar phenomena are expected in our setup, although the energy scales are quite different. Another difference is that there is a sizable matter contribution to the Friedmann equation in our late-time cosmological application. 
In Appendix~\ref{sec: linear analysis}, we first review the linear analysis of the preheating dynamics, and then include the matter component supposing a realistic late-time cosmology.
We found that the resonance structure itself is not significantly changed by the matter component; the matter merely affects the redshift of the oscillation amplitude and the wavenumber. See 
Fig.~\ref{fig: Floquet chart}.
Consequently, the resonance can also occur in the quintessence model as shown in Fig.~\ref{fig: linear power}, which is well consistent with the lattice simulation result (Fig.~\ref{fig: f_1e-3_spectrum}) discussed below.

\bfe{width=0.95\hsize}{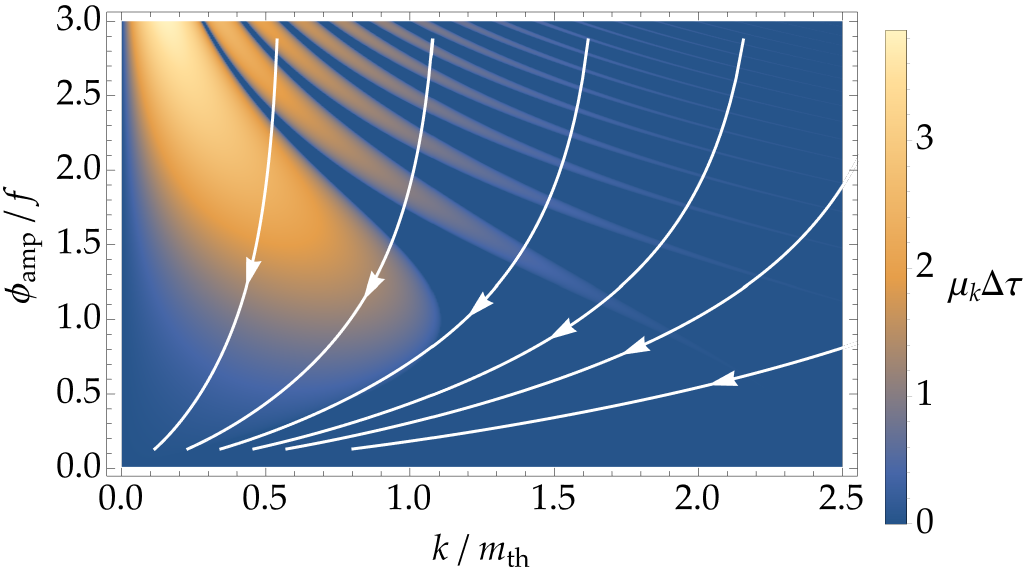}{The growth rate $\mu_k$ (equivalent to $\mu_\kappa$ in Appendix~\ref{sec: linear analysis}) normalized by the oscillation half period $\Delta\tau$ (Eq.~\eqref{eq: delta tau}) as a function of the physical wavenumber $k$ and the oscillation amplitude $\phi_\amp$. White lines indicate the cosmological redshifts of $k$ and $\phi_\amp$ in the expanding universe for $f=10^{-3}\Mpl$,  
$V_0/(3\Omega_\phi H_0^2\Mpl^2)=1.1$,
$H_0=\SI{66}{km/s/Mpc}$, and $\Omega_\um=0.32$ (the quintessence density is given by $\Omega_\phi=1-\Omega_\um$).}{fig: Floquet chart}

\bfe{width=0.95\hsize}{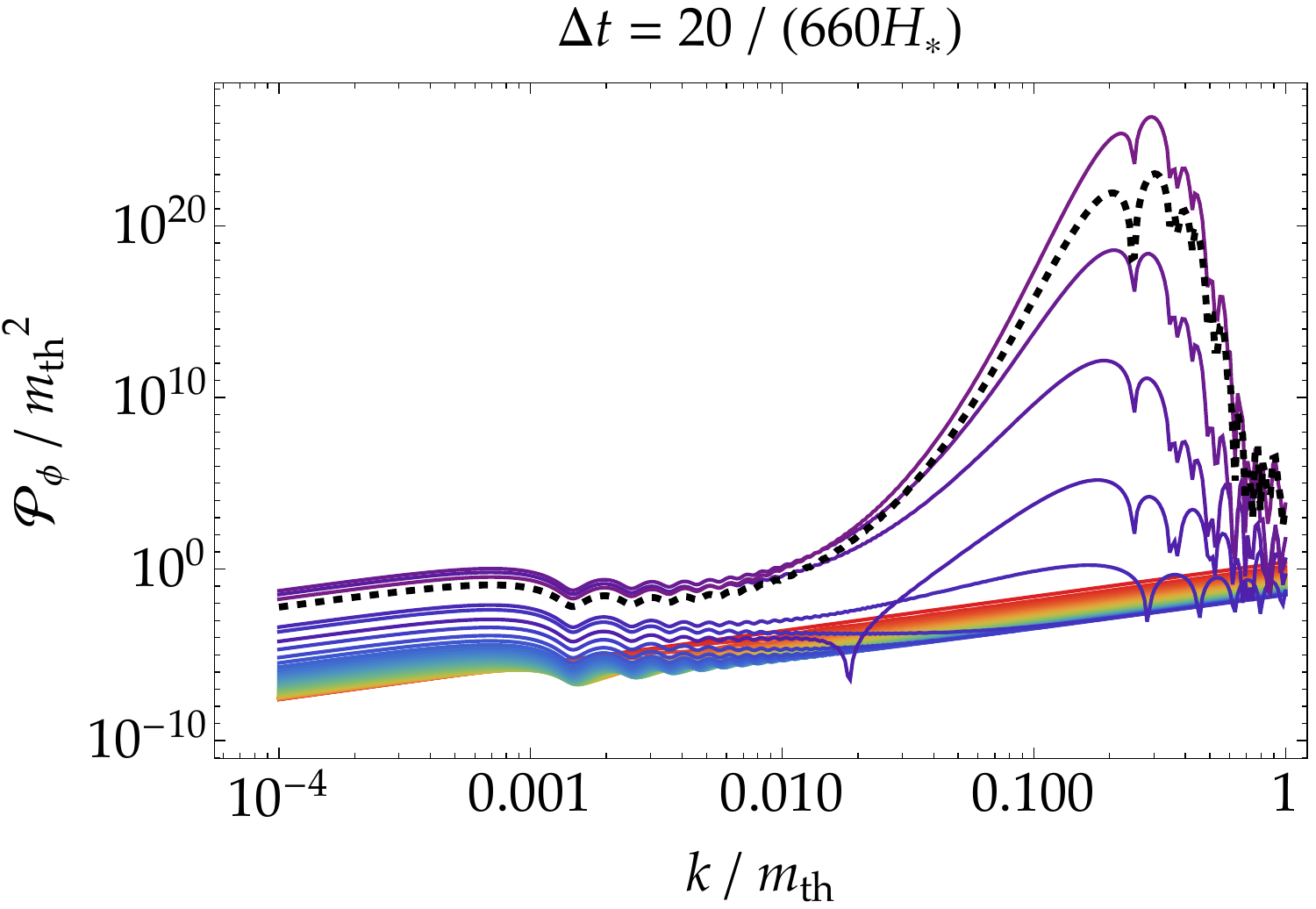}{The evolution of the power spectrum $\calP_\phi=\frac{k^3}{2\pi^2}\abs{\delta\phi_k}^2$ normalised by $m_\uth^2$ from $t=0$ ($z=9$; corresponding to the initial field value, $\phi_\ui/f=7.32$) to $t=900/(660H_*)$ with the time step $\Delta t=20/(660H_*)$ where $H_*=\SI{100}{km/s/Mpc}$ from red to purple. The black dashed line corresponds to today, $t_0=897/(660H_*)$. The evaluated value of $t_0$ and the required initial condition $\phi_\ui$ are slightly different in the lattice simulation, as large fluctuations even change the background expansion history.}{fig: linear power}

The fastest growing mode at each oscillation is phenomenologically found as~\cite{Tomberg:2021bll}
\bae{
    \frac{k_\peak}{am_\uth} \approx \frac{3.54}{\Delta\tau},
    \label{k_peak_tau}
} 
for a sufficiently large oscillation amplitude $|\phi_\amp| \gtrsim 3f$,
where $\Delta \tau$ is the oscillation half period, given by
\bae{\label{eq: delta tau}
    \Delta\tau=\frac{\pi\cosh(\phi_\amp/f)}{\sqrt{2}}.
}
Following the analysis in Refs.~\cite{Karam:2021sno, Tomberg:2021bll} and taking into account the matter component, the initial amplitude can be approximately obtained as follows (see Appendix~\ref{sec: estimate_period})
\bae{
    \cosh\frac{\phi_{\amp,\ui}}{f}\approx\sqrt{\frac{\sqrt{V_0}}{3\sqrt{2}\pi H f}}. \label{cosh_estimate}
}
Combining these, the peak wavenumber can be rephrased in physical units as
\bme{
    k_\peak \approx \SI{4.6e-2}{Mpc^{-1}} \left(\frac{a}{0.9}\right)  \\ 
    \times\left(\frac{f}{10^{-3}\Mpl} \right)^{-1/2}\left( \frac{\tilde{H}}{\SI{63}{km/m/Mpc}} \right), \label{k_peak}
}
where $\tilde{H}\coloneqq\left(\frac{V_0 H^2 }{3 \Mpl^2}\right)^{1/4}$ is the characteristic Hubble scale, evaluated at the onset of the resonance. 
During the tachyonic resonance, the amplitude of $\phi$ decreases, so the oscillation period becomes shorter. Therefore, at each particle production, the characteristic resonant wavenumber scale $k_\text{peak}$ increases.  On the other hand, the estimate in Eq.~\eqref{cosh_estimate} is for the initial oscillation amplitude.

The corresponding frequency $\nu_\peak = k_\peak/(2\pi)$ reads
\begin{align}
    \nu_\peak \approx \SI{
    7.1e-17}{Hz} \left(\frac{k_\peak}{\SI{
    4.6e-2}{Mpc^{-1}}}\right). \label{nu_peak}
\end{align}
Later, we will see that this is the typical frequency of the GWs if the primary peak is dominant. This is also related to the oscillation frequency of $\phi$ at the bottom of the potential as $k_\text{peak}/(a m_\phi) \approx 2.50 / \Delta \tau$.  
The mass of $\phi$ at the bottom of the potential $m_\phi = \sqrt{2} m_\uth$ is given by 
\bme{
    m_\phi = \SI{3.4e-30}{eV} 
    \left(\frac{f}{10^{-3}\Mpl} \right)^{-1} \\
    \times\left(\frac{V_0}{3 \Mpl^2H_0^2}\right)^{1/2}\left(\frac{H_0}{\SI{66}{km/m/Mpc}}\right) . \label{m_phi}
}

As the resonance proceeds, the perturbation is expected to nonlinearly backreact on the background dynamics. In the next subsection, we perform numerical lattice simulations to study the nonlinear dynamics of the quintessence model, including the matter component.

\subsection{Numerical simulation for nonlinear growth of perturbations}\label{sec: lattice}

We use a public package for numerical lattice calculation codes,  $\mathcal{C}\mathrm{osmo}\mathcal{L}\mathrm{attice}$~\cite{Figueroa:2020rrl, Figueroa:2021yhd}, to simulate nonlinear dynamics of the quintessence model discussed in the previous subsection. 

To implement the (dark and baryonic) matter component in the $\mathcal{C}$osmo$\mathcal{L}$attice simulation, we consider a minimally coupled free massive scalar field $\chi$.\footnote{
We have also studied another implementation of dark matter by directly adding a matter component $\rho_\text{m}\propto a^{-3}$ in the Friedmann equations in the $\mathcal{C}$osmo$\mathcal{L}$attice code.  We have observed the consistency of the results at least for a part of the parameter space (the computational cost is high to scan the full parameter space).} 
The details of its properties are not relevant. We just assume that it is sufficiently heavy so that its energy density scales as $\rho_\chi \propto a^{-3}$. Specifically, the Lagrangian density of dark energy and matter is 
\beae{
    &\mathcal{L}= - \frac{1}{2}g^{\mu\nu}\partial_\mu \phi \partial_\nu \phi - \frac{1}{2} g^{\mu\nu}\partial_\mu \chi \partial_\nu \chi - V(\phi, \chi), \\
    &V(\phi, \chi)= V_0 \tanh^{2} \left( \frac{\phi}{f} \right) + \frac{1}{2} m_\chi^2 \chi^2,
}
where $m_\chi$ is the mass of the matter component $\chi$. 
In the $\mathcal{C}$osmo$\mathcal{L}$attice simulation, perturbations of the metric are not taken into account, and only the scale factor $a(t)$ is kept track of.

We consider six sets of parameters for the lattice simulation (see Table~\ref{tab: params_for_lattice}).  The free parameter in this analysis is $f$, while the other parameters are fixed to match the MCMC results, \textit{cf.}, Table \ref{tab:MCMC_TDE}. Specifically, we adopt $H_0 = 66\,\mathrm{km}/\mathrm{s}/\mathrm{Mpc}$, $\Omega_\mathrm{m} = 0.32$, and $a_\mathrm{t} = 0.9$ as a benchmark point close to the maximum a posteriori in the MCMC results. We set the lattice grid number $512$ and conducted three-dimensional lattice simulations from $a = 0.1$.\footnote{In some cases, we set the grid number to $128$ to reduce the computational cost. See below for more details.}

\begin{table}
\renewcommand{\arraystretch}{1.3}
 \begin{center}
  \caption{Six parameter sets for lattice simulation. $\phi_\ui$ is the scalar field value at the initial time $a = 0.1$.  \label{tab: params_for_lattice}}
  \begin{ruledtabular}
  \begin{tabular}{lccccccc}
    Parameters & \quad & 1 & 2 & 3 & 4 & 5 & 6 \\
    \hline
    $f \ [\Mpl]$ & & $10^{-4}$ & $10^{-3.5}$ & $10^{-3.25}$&$10^{-3}$ &$10^{-2.5}$ &$10^{-2}$ \\
    $ V_0 \ [3\Omega_\phi H^2\Mpl^2]$ & & $ 1.46$ & $ 1.52$& $ 1.45$&$ 1.1$& $ 1.13$& $ 1.13$ \\
    $\phi_\mathrm{i} \ [f]$ & & $ 9.68$ & $ 8.53$ &$ 7.92$& $ 7.33$ & $ 6.19$ & $ 5.05$ \\
  \end{tabular}
  \end{ruledtabular}
 \end{center}
\end{table}

The initial conditions are as follows. We take the initial velocity of $\phi$ to be vanishing. As we will see, perturbations grow exponentially, so quantities such as the time when backreaction becomes relevant are only logarithmically sensitive to the initial condition.  For simplicity, we take the Bunch--Davies initial condition. More precisely, we neglect the (small) tachyonic mass in the Banch--Davies condition, which is also the default prescription in $\mathcal{C}$osmo$\mathcal{L}$attice. We take the periodic boundary condition.

\subsubsection{Approximate realisation of TDE}

An important effect in the nonlinear regime is backreaction from the produced perturbations to the homogeneous background. When the exponentially increasing energy density of the perturbations becomes comparable with the background, the notion of the homogeneous background ceases to be effective.  Then, the homogeneous dark energy field fragments into inhomogeneous pieces. In this process, a significant fraction of the energy density is transferred from the background to perturbations, so the background amplitude significantly reduces when the backreaction becomes significant.  Subsequently, the background only probes the quadratic part of the potential, so the preheating-like dynamics are shut down. With all these effects, the background dynamics are changed from those in the linear perturbation analyses (and of course also from the background-only analyses). 
Figure~\ref{fig: f_1e-3_scal} shows the dynamics of the simulation-box average $\expval{\phi}$ computed in the lattice simulation for the case $f = 10^{-3} \Mpl$ as an example. We start our lattice simulation at scale factor $a = 0.1$ with $\tilde t = 0$, where $\tilde t$ is physical time normalised by $660H_* = 1000H_0$.
For comparison, we also plot the dynamics of the homogeneous mode without the backreaction with the yellow dotted line. After several oscillations, the two lines, initially overlapping with each other, become completely different due to the backreaction from particle production.
For the case with $f = 10^{-3} M_\mathrm{Pl}$, the nonlinear effects remain negligible at $a = 1$.

\bfe{width=0.95\hsize}{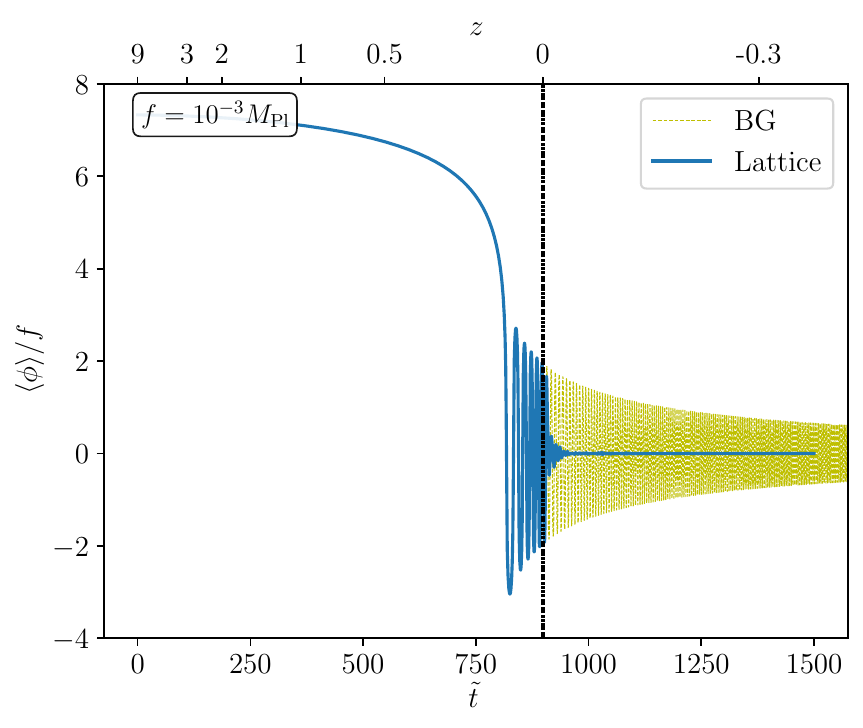}
{Dynamics of the simulation-box average field value $\expval{\phi} /f $ with (lattice simulation; blue solid line) and without (yellow dotted line) the  backreaction from perturbations for $f = 10^{-3} \Mpl$. $\tilde t$ is physical time normalized by $660H_* = 1000H_0$. The vertical black dotted line represents the current time $a = 1$. }{fig: f_1e-3_scal}

\begin{figure*}
	\centering
	\begin{tabular}{c}
		\begin{minipage}{0.48\hsize}
			\centering
			\includegraphics[width=\hsize]{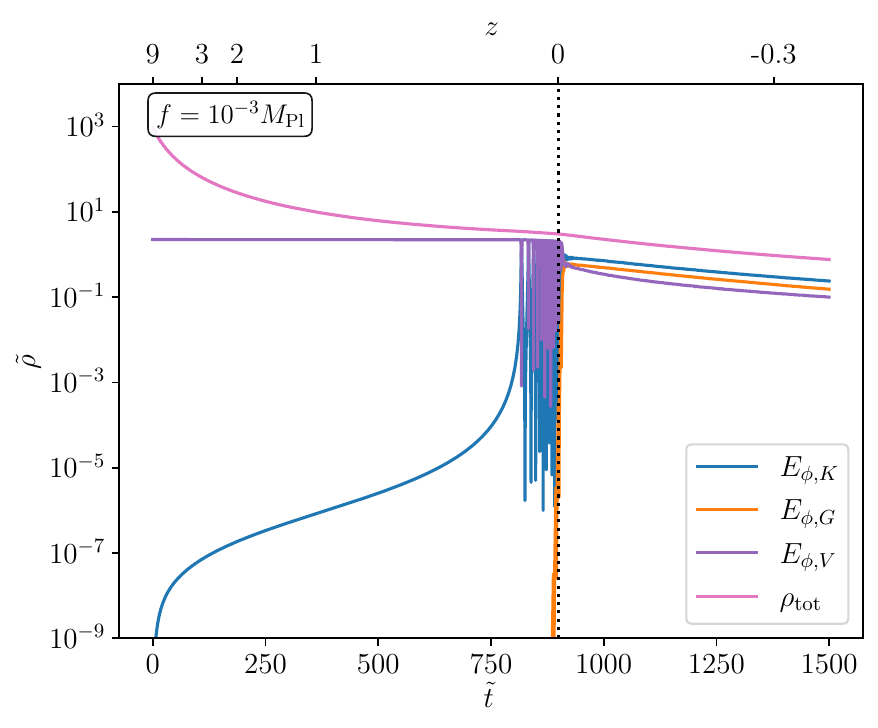}			
		\end{minipage} 
		\begin{minipage}{0.48\hsize}
			\centering
			\includegraphics[width=\hsize]{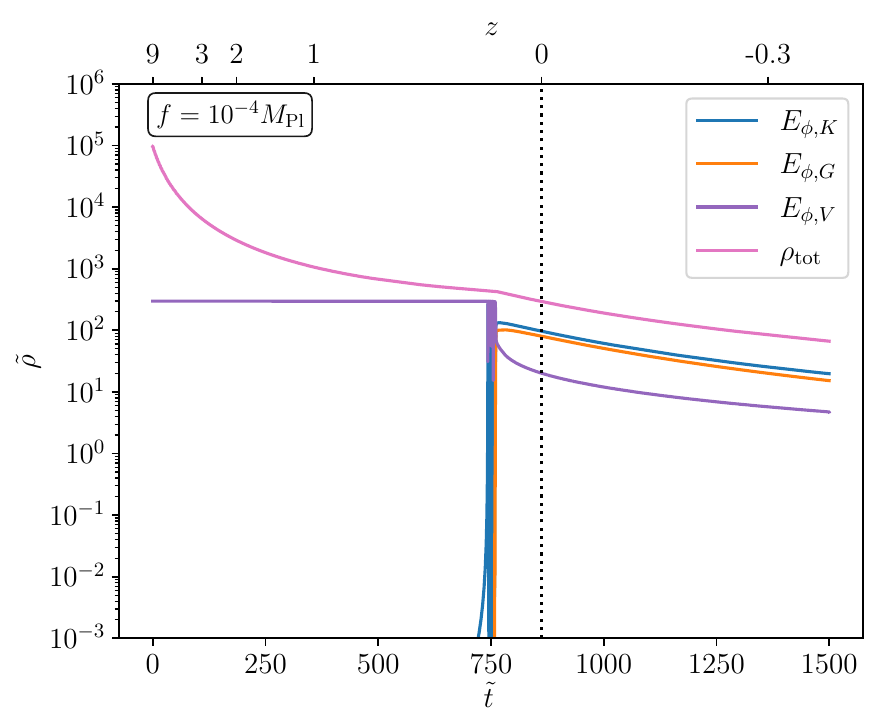}
		\end{minipage}
	\end{tabular}
	\caption{Time dependence of the spatially averaged energy density components of scalar field: the kinetic energy $E_{\phi, K}$ (blue), the gradient energy $E_{\phi, G}$ (orange), the potential energy $E_{\phi, V}$ (purple), and the total energy $\rho_{\text{tot}}$ (pink) for $f=10^{-3}\Mpl$ (left) and $f=10^{-4}\Mpl$ (right). $\tilde t$ is physical time normalized by $660H_* = 1000H_0$. $\tilde \rho$ is energy density normalized by $(f\times660H_*)^2$. The vertical black dotted line represents the current time $a = 1$.  
    The output cosmological parameters 
    are found as $H_0/(\mathrm{km/s/Mpc}) = 66.3$ and $\Omega_\um = 0.316$ for $f=10^{-3}\Mpl$, while $H_0/(\mathrm{km/s/Mpc}) = 65.1$ and $\Omega_\um = 0.327$ for $f=10^{-4}\Mpl$. 
    }
	\label{fig: lattice_energy}
\end{figure*}

\begin{figure*}
	\centering
	\begin{tabular}{c}
		\begin{minipage}{0.48\hsize}
			\centering
			\includegraphics[width=\hsize]{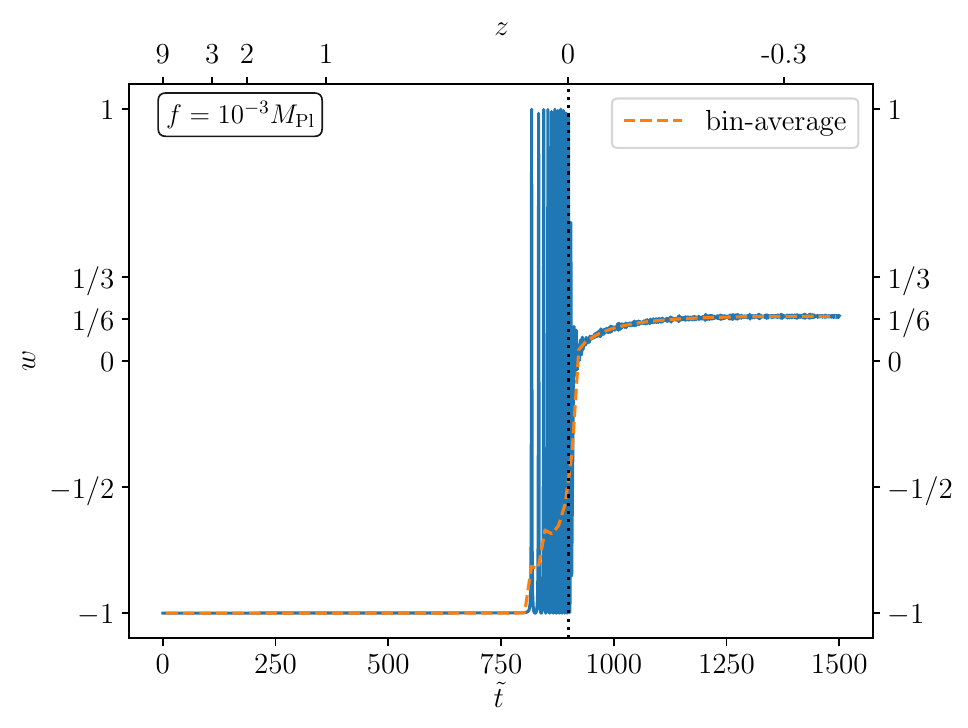}			
		\end{minipage} 
		\begin{minipage}{0.48\hsize}
			\centering
			\includegraphics[width=\hsize]{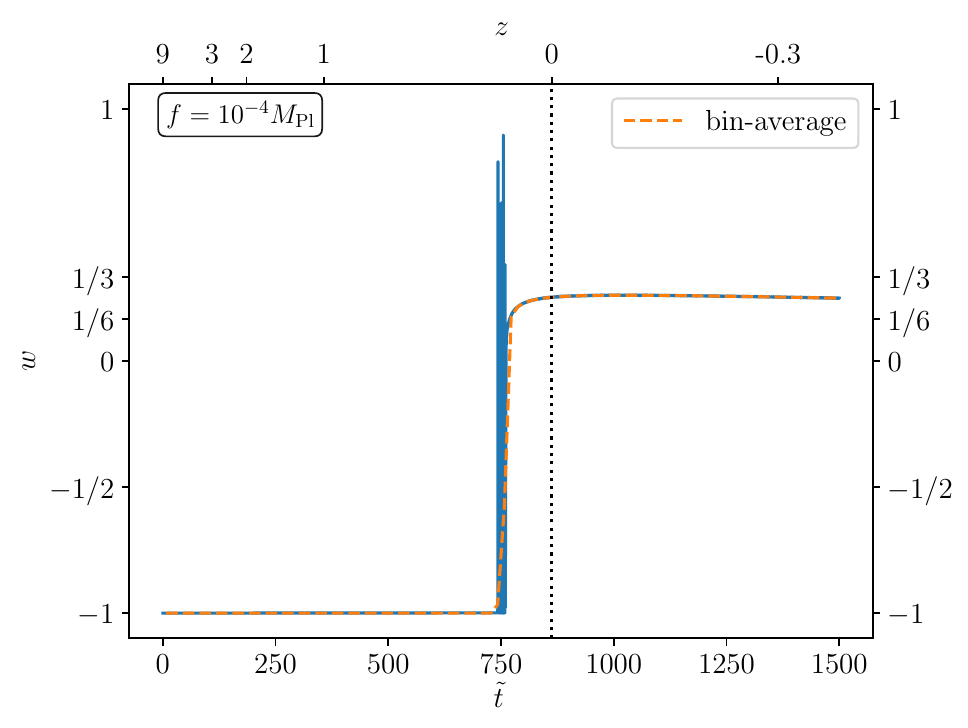}
		\end{minipage}
	\end{tabular}
	\caption{Time dependence of the \ac{EoS} parameter of the scalar field $\phi$ (solid blue line) for $f=10^{-3}\Mpl$ (left) and $10^{-4}\Mpl$ (right). $\tilde t$ is physical time normalized by $660H_* = 1000H_0$. The dashed orange line is the result of the time average.  The vertical black dotted line represents the current time $a = 1$.  
    }
	\label{fig: lattice_EoS}
\end{figure*}

\begin{figure*}
	\centering
	\begin{tabular}{c}
		\begin{minipage}{0.48\hsize}
			\centering
			\includegraphics[width=\hsize]{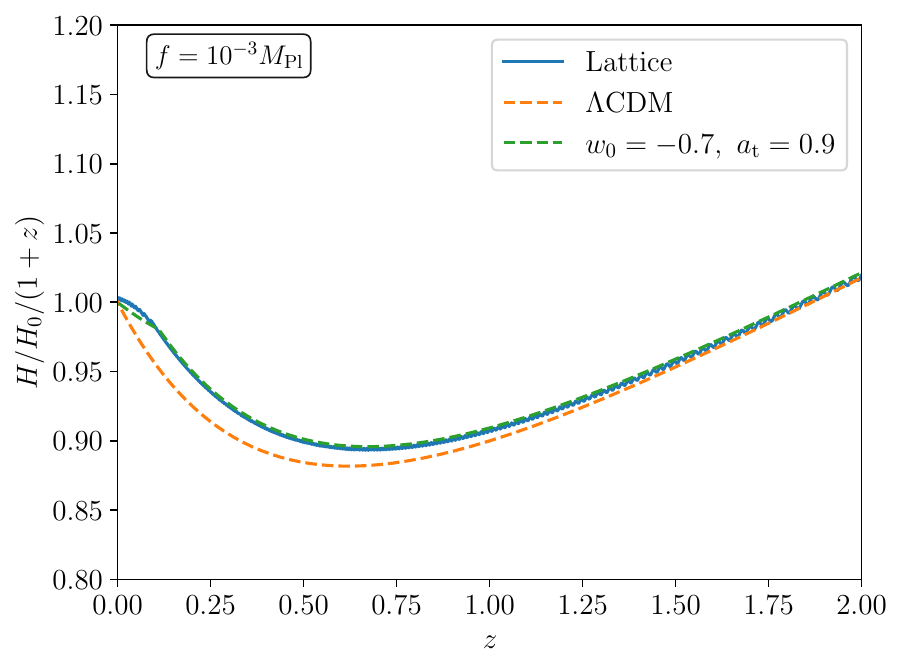}			
		\end{minipage} 
		\begin{minipage}{0.48\hsize}
			\centering
			\includegraphics[width=\hsize]{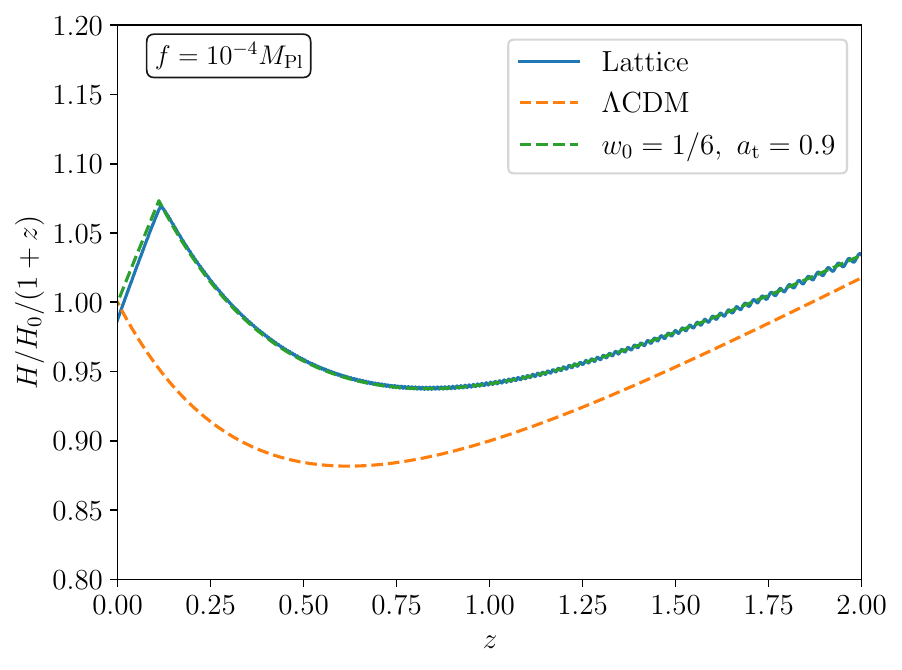}
		\end{minipage}
	\end{tabular}
	\caption{The lattice results on the evolution of the Hubble parameter (solid blue line) for
    $f = 10^{-3} \Mpl$ (left) and $f = 10^{-4}\Mpl$ (right). 
    They are well fit by the \ac{TDE} model shown by the red dashed lines.
    }
	\label{fig: lattice_hubble}
\end{figure*}

Various background quantities obtained from the lattice simulation are shown in Figs.~\ref{fig: lattice_energy}, \ref{fig: lattice_EoS}, and \ref{fig: lattice_hubble} for two parameter choices $f/\Mpl = 10^{-3}$ and $10^{-4}$. 
Figure~\ref{fig: lattice_energy} shows the time dependence of the spatially averaged total energy density $\rho_\text{tot}$ as well as spatially averaged energy-density components: kinetic energy $E_{\phi,K}$, gradient energy $E_{\phi,G}$, and potential energy $E_{\phi,V}$.  By comparing the left and right panels, one can see that the intermediate (tachyonic resonance) dynamics is indeed shorter for a smaller value of $f$. 

Figure~\ref{fig: lattice_EoS} shows the time dependence of the \ac{EoS} parameter. We also plot the time-averaged \ac{EoS} parameter with the averaging window $\Delta \tilde t = 15$,
which would be more relevant for comparisons with observations. In the left panel, $w$ suddenly increases at around $\tilde t \approx 800$ from $-1$ toward $\approx -0.5$ at present, denoted by the vertical black dotted line. This roughly reproduces the behaviour of the broken-linear model~\eqref{broken-linear_model}. After that, $w$ further increases and reaches an approximate plateau. In this larger time scale, the behaviour can be interpreted as the step model~\eqref{eq: TDE_EoS}. In the right panel, with the smaller value of $f$, the transitional behaviour is sharper, so the model is well approximated by the step-like TDE model~\eqref{eq: TDE_EoS}.

A rough estimate of the value of the \ac{EoS} parameter after particle production can be given as follows. After the end of the tachyonic resonance, the amplitude of $\phi$ decreases significantly, so the quadratic mass term dominates the potential, and the self-couplings can be neglected. Let us consider two extreme limits that serve as upper and lower bounds on $w$. On one hand, if we neglect the produced particles, the energy density and pressure are carried by the residual homogeneous condensate of $\phi$, which has $w = 0$. 
On the other hand, if we neglect the condensate of $\phi$ and view the post-resonance state as relativistic $\phi$ particles, they have $w = 1/3$. Tachyonic resonance produces semirelativistic particles, so the fully relativistic limit would be an overestimate of the \ac{EoS} parameter of the $\phi$'s fragments. In addition, we generally expect that the final state is a mixture of the residual homogeneous condensate of $\phi$ and the (semi)relativistic perturbations of $\phi$.  Therefore, we generally expect $0 \lesssim w \lesssim 1/3$ just after the end of the tachyonic resonance,\footnote{
Since the tachyonic resonance ends when the backreaction becomes significant, \ie, when the background and fluctuations become comparable, one would naively expect an intermediate situation of these two limits as a $1:1$ mixture of nonrelativistic matter and radiation, which has $w = 1/6$. This value matches the lattice for $f=10^{-3} \Mpl$ result well (the left panel of Fig.~\ref{fig: lattice_EoS}), but it is accidental because it does not match the lattice result for $f=10^{-4} \Mpl$ (the right panel of Fig.~\ref{fig: lattice_EoS}) and also because the ratio among the kinetic, gradient, and potential energy-density components of $\phi$ is not quantitatively consistent with the above interpretation of the $1:1$ mixture of nonrelativistic matter and radiation. 
} 
which is consistent with the numerical lattice result shown in Fig.~\ref{fig: lattice_EoS}. 

As the Universe continues to expand, relativistic modes redshift faster than nonrelativistic matter.  
Hence, the \ac{EoS} parameter in our model finally asymptotes to the matter-dominated value, $w=0$.
In this phase, the Universe is dominated by $\chi$, representing both dark and baryonic matter, and by $\phi$, which used to be dark energy but has transmuted into the second component of dark matter. Of course, this final value depends on our choice of the vanishing cosmological constant. If we add a positive or negative cosmological constant, the future asymptotic value of $w$ is $-1$ or $+1$, respectively.\footnote{
With a negative cosmological constant, the Hubble expansion turns into contraction, and the energy density is finally dominated by the kinetic energy of $\phi$~\cite{Linde:2001ae, Felder:2002jk}.
} A sizable cosmological constant can also affect the value of $w$ during and right after the end of the resonance.  Let us also comment that $w$ could transiently take a larger value for a larger power $n$ in the potential~\eqref{quintessence_model} even without the cosmological constant (see also footnote~\ref{footnote: w in n}). 

Figure~\ref{fig: lattice_hubble} shows the redshift-dependence of the Hubble parameter.  One can see that the behaviour of the quintessence model can be well approximated by the step-like TDE model. 
The parameter values of the step model, $w_0$ and $a_\text{t}$, in the figure are not the best-fit values but round values that approximately fit the lattice results. 

We have shown a successful realisation of the \ac{TDE} model by a scalar quintessence.
While the transitioning time is understood well by the homogenous approximation, 
the dynamics of the scalar field deviate after several oscillations due to nonlinear effects, which determine the end of the resonance and the subsequent \ac{EoS} value.

\subsubsection{Scalar Perturbations}

\bfe{width=0.95\hsize}{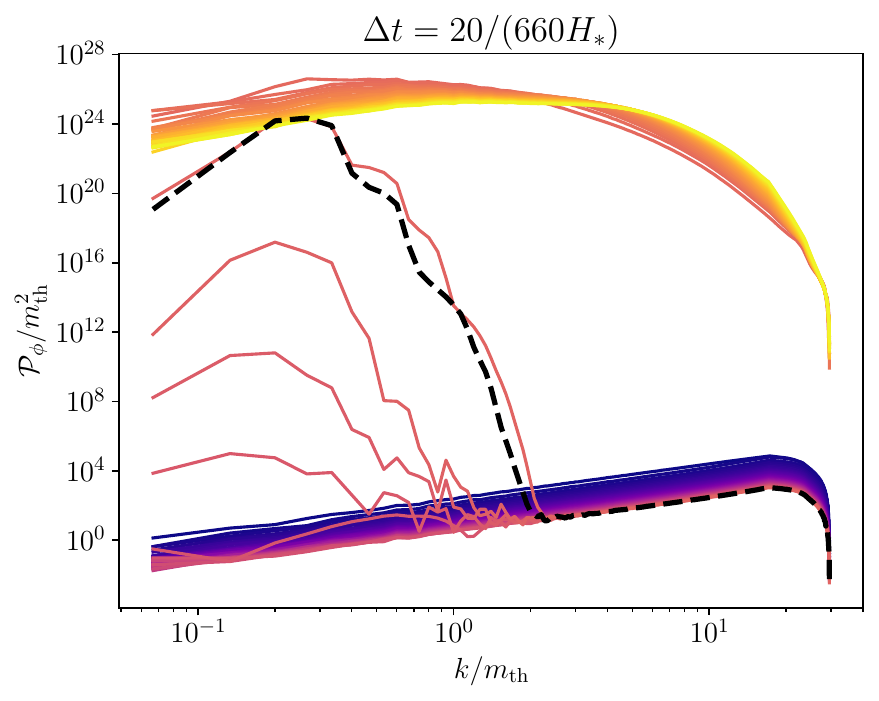}{The time evolution of the scalar power spectrum for $f = 10^{-3} \Mpl$ from $t = 0$ ($z = 9$) with the time step $\Delta t = 20/(660H_*)$ from blue to yellow. See Fig.~\ref{fig: Floquet chart} for the counterpart in the linear analysis. The black dashed line corresponds to today $t_0=899/(660H_*)$.}{fig: f_1e-3_spectrum}

\begin{figure*}
	\centering
	\includegraphics[width=0.9\hsize]{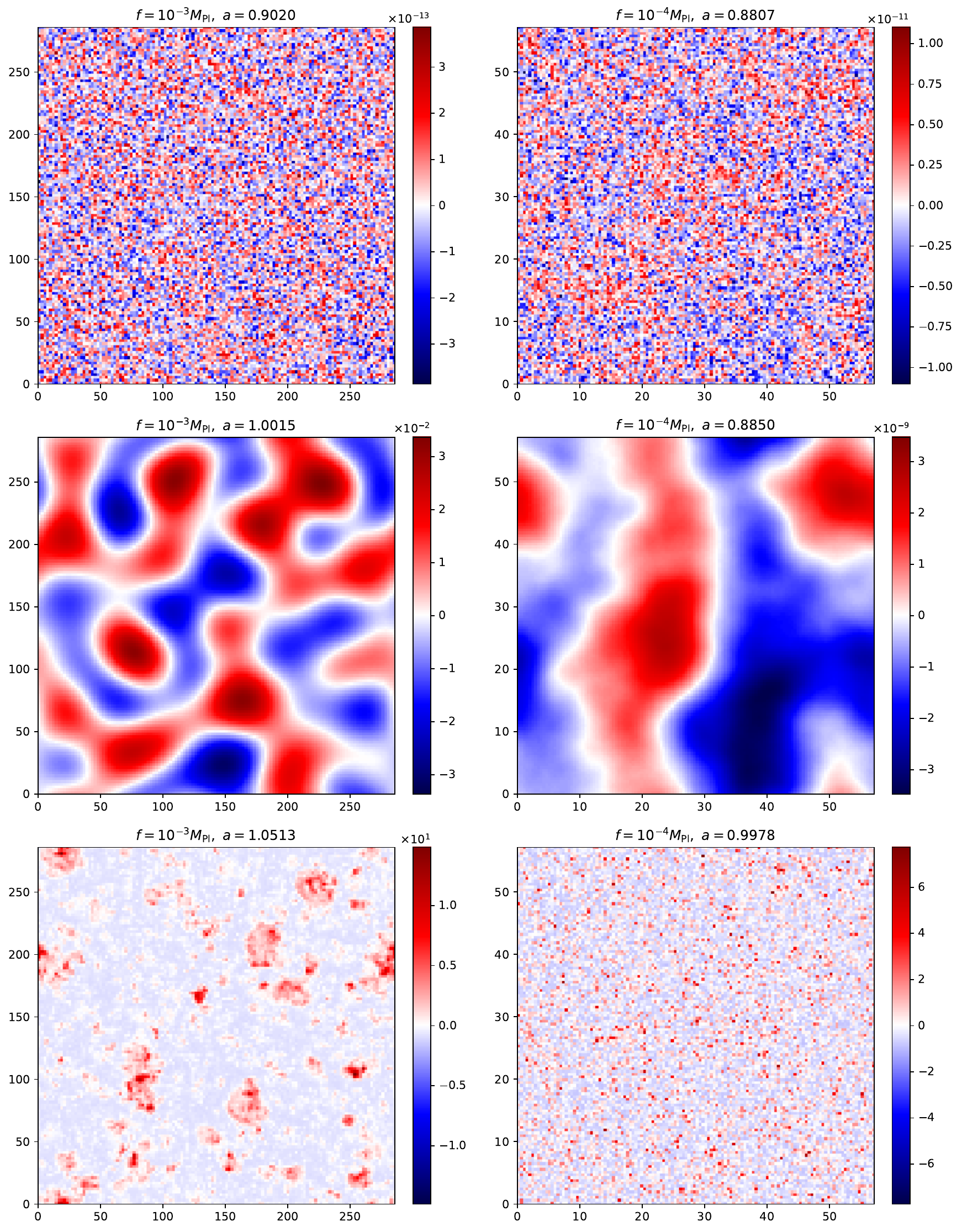}
	\caption{The time evolution of dark energy density contrast $\delta_\phi=(\rho_\phi-\expval{\rho_\phi})/\expval{\rho_\phi}$, where $\expval{\rho_\phi}$ is the simulation box average of dark energy, on a two-dimensional spatial slice for $f=10^{-3}\Mpl$ (left) and $f=10^{-4}\Mpl$ (right). 
    The horizontal and vertical axes show physical distance in units of $\si{Mpc}$.  
    The top, middle, and bottom rows correspond to the early, intermediate, and late times. 
	Here, we set the grid number to $128$.}
	\label{fig: energy_snapshot}
\end{figure*}

\bfe{width=0.95\hsize}{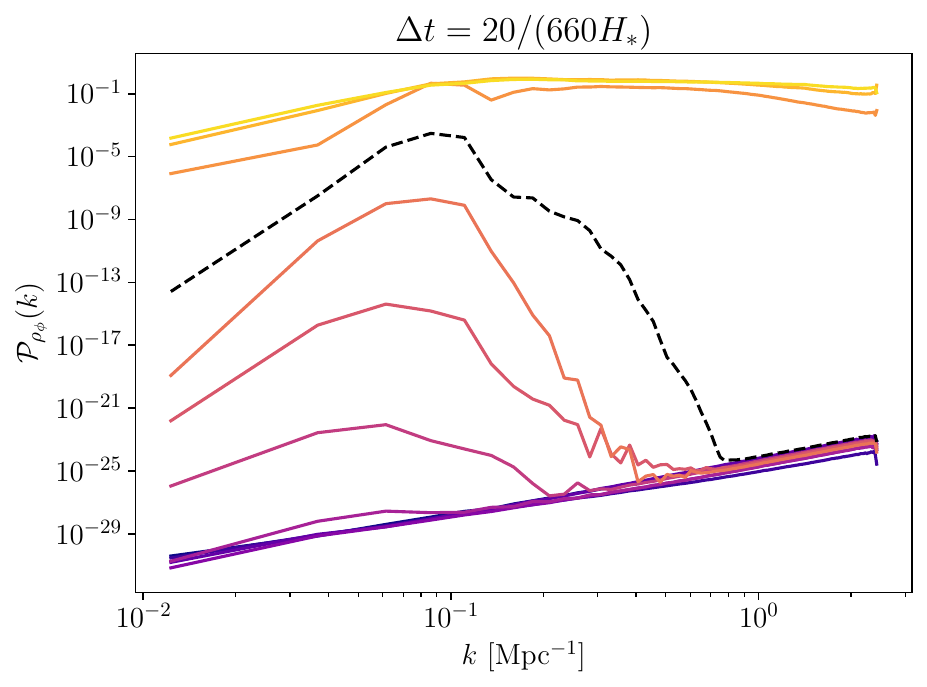}{Time evolution of the dimensionless power spectrum of the dark energy density for $f = 10^{-3} \Mpl$ with the time step $\Delta t = 20/(660H_*)$ from blue to yellow. Here, we set the grid number to $128$. The black dashed line approximately represents today, $a= 1.0015$, corresponding to the middle left panel in Fig.~\ref{fig: energy_snapshot}. }{fig: f_1e-3_Enpower}

We then investigate the scalar fluctuations in more detail.
Figure~\ref{fig: f_1e-3_spectrum} shows the lattice result on the evolution of the scalar power spectrum $\calP_\phi(k)$. 
The overall behaviour is quite similar to its linear counterpart shown in Fig.~\ref{fig: linear power}, particularly in terms of the peak position, growth rate, and amplitude of the perturbations. However, shortly after the current time $t_0$, the growth of the peaked perturbations saturates, and their energy is transferred to other modes via nonlinear mode-coupling effects.

The peak wavenumber is given by Eq.~\eqref{k_peak_tau}, reading $k_\peak \sim 0.161m_\uth \sim \SI{0.053}{Mpc^{-1}}$ for $f = 10^{-3}\Mpl$, where $\phi_{\amp, \ui} / f =2.88$ and $\Delta\tau(\phi_{\amp,\ui} / f) = 19.8$. This field value is defined as the mean of the first peak and trough amplitudes obtained from the scalar field dynamics in the lattice simulation, as shown in Fig.~\ref{fig: f_1e-3_scal}. 
On the other hand, we can estimate $k_\text{peak}$ using Eq.~\eqref{k_peak}. In this case, the value of $\tilde H$ is about $\sim \SI{63}{km/s/Mpc}$ with $f = 10^{-3}\Mpl$, so $k_\text{peak} \sim \SI{0.046}{Mpc^{-1}}$. These two values of $k_\text{peak}$ are consistent within a $10\%$ error.
Note that the actual peak of the spectrum after the end of tachyonic resonance is different from these values because (1) the resonance wavenumber grows during the tachyonic resonance process, and (2) the second and higher instability bands may also affect the spectral shape.

Figure~\ref{fig: energy_snapshot} shows snapshots of dark energy density distribution. 
The top panels show an early stage of the tachyonic resonance. The significant growth of perturbations at a characteristic scale $k_\peak$, which depends on $f$, is visible in the middle panels. Later, the coupling among modes is relevant, and the power is transferred to shorter scales (see the bottom panels). In the left panel ($f=10^{-3}\, \Mpl$), one can see clumps of overdense regions shown in red. This may show the formation of a long-lived nontopological soliton called an oscillon/I-ball~\cite{Bogolyubsky:1976yu, Gleiser:1993pt, Copeland:1995fq, Kasuya:2002zs}.  A much longer simulation will be required to draw a more robust conclusion, which we leave for future work.

Figure~\ref{fig: f_1e-3_Enpower} shows the evolution of the dimensionless power spectrum of dark energy density perturbation, $\mathcal P_{\rho_\phi} (k)= \frac{k^3}{2\pi^2}\abs{\delta_{\phi, k}}^2$, where $\delta_{\phi, k}$ represents the Fourier mode of dark energy density contrast $\delta_\phi=(\rho_\phi-\expval{\rho_\phi})/\expval{\rho_\phi}$ shown in Fig.~\ref{fig: energy_snapshot}.
In Sec.~\ref{sec: ISW}, we use this power spectrum to calculate the gravitational potential appearing in the Poisson equation and then obtain the \ac{CMB} anisotropies.

As we have seen so far, the rolling dynamics of the quintessence field $\phi$ are rather suddenly accelerated around the edge of the plateau if $f \ll \Mpl$.  Compared to the regular thawing quintessence scenario, our quintessence scenario involves \emph{rapid thawing}. Shortly after the onset of rapid rolling, the tachyonic resonance is developed, and the inhomogeneity grows. As long as the oscillation amplitude of the field is large enough to probe the non-quadratic part of the potential, the inhomogeneity developed on some scale is transferred to other scales via nonlinear interactions, so the momentum distribution approaches that of the thermal one. Thus, the rapid thawing is followed by the thermalisation process, which may or may not be completed.  It is tempting to dub the quintessence model as \emph{meltdown quintessence}.  Note, however, that Fourier modes decouple once the oscillation amplitude of the background is sufficiently reduced, so the system is not in thermal equilibrium at late time.

\subsubsection{Gravitational Waves}

The scalar resonance results in a large anisotropic stress in 
their energy-momentum tensor, which yields stochastic generation of \acp{GW}.
The \ac{GW} \ac{EoM} is given by
\bae{
\ddot h_{ij} + 3H\dot h_{ij} -\frac{\nabla^2}{a^2} h_{ij} = \frac{2}{a^2 M_\mathrm{P}^2} \Pi_{ij}^\mathrm{TT},
}
where $h_{ij}$ is the transverse-traceless tensor perturbation of the spacetime metric, and $\Pi_{ij}^\mathrm{TT}$ is the transverse-traceless part of the anisotropic stress. 
We can directly get the evolution of the \ac{GW} density parameter $\Omega_\mathrm{GW}$ from $\mathcal C \mathrm{osmo} \mathcal L \mathrm{attice}$~\cite{CosmoLatticeGW}, which implements this \ac{EoM}. 

\bfe{width=0.95\hsize}{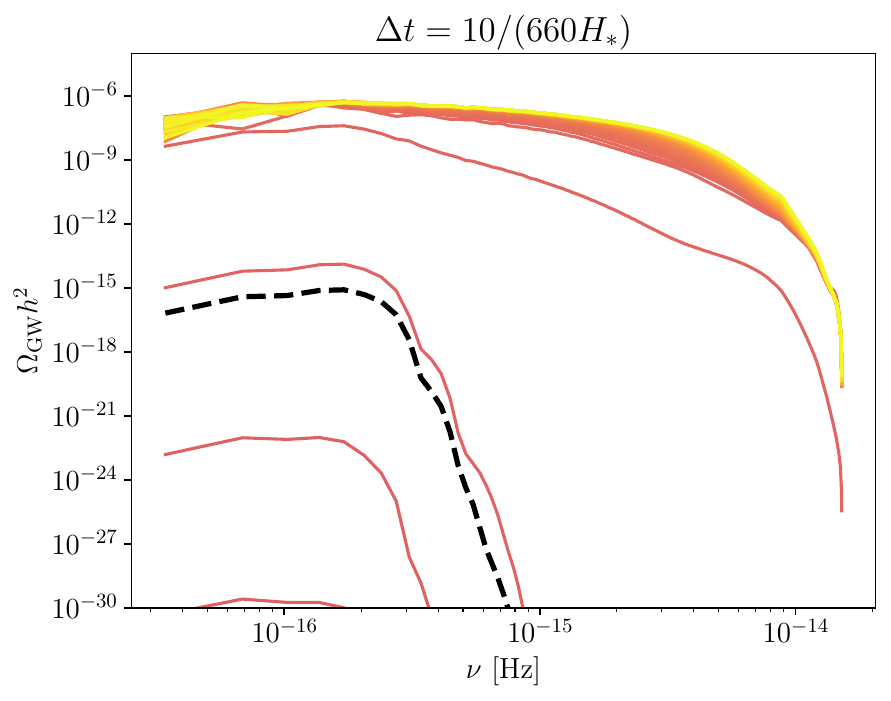}{The time evolution of the \ac{GW} spectrum $\Omega_\GW(\nu)h^2$ defined by Eq.~\eqref{eq: OmegaGW h2} for $f = 10^{-3} \Mpl$ with the time step $\Delta t = 10/(660H_*)$ from red to yellow. The dashed black line corresponds to today $t_0=899/(660H_*)$.}{fig: GW_time_evol}

Figure~\ref{fig: GW_time_evol} shows the time evolution of the \ac{GW} spectrum $\Omega_\GW(\nu)$ defined by 
\bae{\label{eq: OmegaGW h2}
    \Omega_\GW h^2=\int\Omega_\GW(\nu)h^2\dd{\ln{\nu}},
}
for $f = 10^{-3} \Mpl$. 
One finds that the \ac{GW} spectrum also grows exponentially, and shortly after the current time, it saturates. 
Though it is not strongly peaked compared to the scalar perturbations, the characteristic frequency $\nu_\text{peak}$ of the produced \acp{GW} is indeed related to the scalar peak mode~\eqref{nu_peak}:
\bae{
 \nu_\mathrm{peak} = \frac{1}{2\pi}k_\mathrm{peak} \sim 7 \times 10^{-17} \, \mathrm{Hz} \left(\frac{10^{-3}\Mpl}{f}\right)^{1/2}.
}
Note that the actual peak frequency deviates from $\nu_\text{peak}$ for $f \ll 10^{-4} \Mpl$~\cite{Tomberg:2021bll}. We will come back to this point later. 

In Figs.~\ref{fig: f_1e-3_spectrum}, \ref{fig: f_1e-3_Enpower}, and \ref{fig: GW_time_evol}, power can be transferred from the peak scale to higher frequencies. Around the highest frequency (beyond the Nyquist frequency) in each case, the decrease (if any) of power is considered to be affected by the resolution limit of the lattice. 
This point is particularly relevant when we discuss the observational prospects of the GWs in the next subsection.

\subsection{Observational Implications}

\subsubsection{\Acl{ISW} effect on \ac{CMB}}\label{sec: ISW}

The inhomogeneity in dark energy would directly be observed through \ac{CMB} anisotropies.
Its time evolution causes the time-dependent gravitational potential, which affects the \ac{CMB} photon propagation. 
This effect is known as the \ac{ISW} effect~\cite{Seljak:1996is}.

The source function of the temperature perturbations from the \ac{LISW} effect is 
\begin{equation}
    S^{\mathrm{LISW}}_T(k, \tau)  = e^{-\kappa}(\dot\Phi - \dot \Psi) \sim 2\dot \Phi, \label{eq:LISW_source}
\end{equation}
where $\Phi$ and $\Psi$ are the Bardeen potentials, and $\kappa$ is the optical depth, which we neglect as we consider the late time $z < 10$. For simplicity, we neglect anisotropic stress, so $\Phi + \Psi = 0$.\footnote{The second order perturbations in the energy-momentum tensor contribute to the anisotropic stress, which is the source of the \acp{GW}.  In the parameter space where the dark-energy contribution to the \ac{ISW} effect becomes comparable to the standard \ac{ISW} effect, the perturbative expansion becomes questionable. For our purposes, an order-of-magnitude estimate is sufficient, and a more precise study is left for future work.}  

The gravitational potential $\Phi$, appearing in the \ac{LISW} source function~\eqref{eq:LISW_source}, is determined by the Poisson equation:
\begin{equation}
    \label{eq: poisson}
    k^2\Phi(k,\tau) = 4\pi Ga^2(t)\expval{\rho}(t)\delta(k, t),
\end{equation}
where $\expval{\rho}$ is the spatial average of the total energy density, and $\delta=(\rho-\expval{\rho})/\expval{\rho}$ is its perturbation.
During the matter-dominanted era, $\expval\rho \propto a^{-3}$ and $\delta\propto a$, so $\Phi = \mathrm{const}$. The ISW effect is not effective in the matter-dominated Universe since the source term vanishes.  However, $\delta$ and hence $\Phi$ rapidly grow during the resonant phase. We substitute the numerically obtained dark energy density into Eq.~\eqref{eq: poisson} and obtain $\Phi$ and hence $S^{\mathrm{LISW}}_T(k, \tau)$. 

The spherical harmonic expansion of the temperature fluctuations is given by
\begin{equation}
    \Delta^{\mathrm{LISW}}_{T,\ell}(k, \tau = \tau_0) = \int^{\tau_0}_{\tau_\mathrm s} \dd{\tau} S^{\mathrm{LISW}}_T(k, \tau)j_\ell[k(\tau_0 - \tau)],
\end{equation}
where $\ell$ denotes the multipole, and $j_\ell(x)$ is the spherical Bessel function. 
The angular power spectrum of the temperature anisotropy from \ac{LISW} hence reads
\begin{equation}
    C_\ell^{\mathrm{LISW}} = (4\pi)^2\int k^2 \dd{k} \abs{\Delta^{\mathrm{LISW}}_{T,\ell}(k, \tau = \tau_0)}^2. 
\end{equation}

Combining these equations, we obtain the \ac{CMB} temperature anisotropy from dark energy perturbations. In our analysis, dark matter and dark energy densities are uncorrelated. Since dark energy perturbations are random, we can get the full anisotropies $C_\ell^{\mathrm{tot}}$ by simply adding two terms,
\bae{
    C_\ell^{\mathrm{tot}} = C_\ell^{\mathrm{LISW,DE}} + C_\ell^{\mathrm{BG}},
    \label{eq: Cl_tot}
}
where $C_\ell^{\mathrm{LISW,DE}}$ denotes the temperature anisotropy from dark energy perturbations calculated from lattice simulations and $C_\ell^{\mathrm{BG}}$ is the background temperature anisotropy in the \ac{TDE} background (without perturbations of dark energy) calculated by our modified \texttt{CAMB} used in Sec.~\ref{sec: MCMC analysis}.

\bfe{width=0.95\hsize}{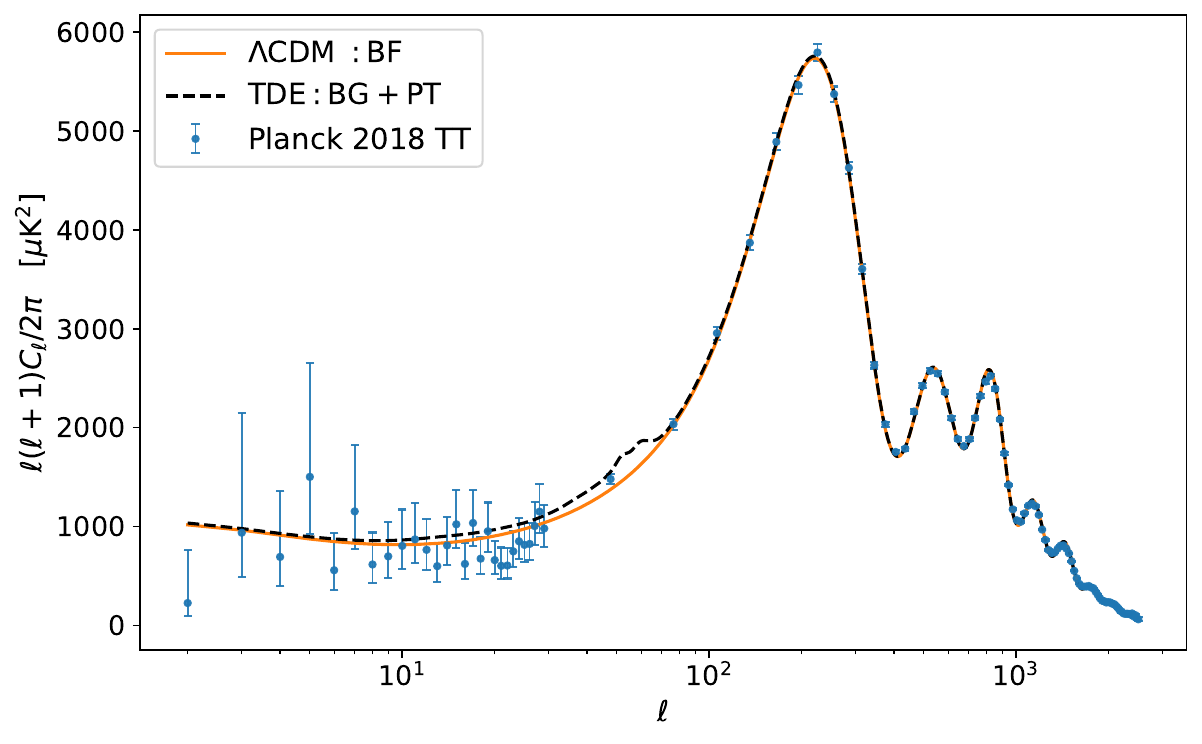}{Effects of dark energy perturbations on the CMB temperature angular power spectrum via the ISW effect. The blue points with error bars represent the Planck data~\cite{Planck:2018vyg}. The solid orange curve shows the standard $\Lambda$CDM prediction with the Planck best-fit parameters. The dashed black line includes the contribution from the late-time ISW effect caused by the growth of the gravitational potential induced by the tachyonic resonance obtained from a lattice simulation with a grid number of 128. The decay constant value is $f = 10^{-3.25}  \Mpl$, around which the effect is maximal.
This model is then distinguishable at around $1\sigma$ level as indicated by the chi-squared~\eqref{eq: chi2}.
}{fig: ISW_f_1e-325}

Figure~\ref{fig: ISW_f_1e-325} shows the ISW effect obtained from our lattice simulation for $f = 10^{-3.25} \Mpl$, together with the Planck results~\cite{Planck:2018vyg}. 
The ISW effect is almost maximal for this value of the decay constant $f$, 
leading to a deviation from the Planck data around $\ell\sim50$ at the several-sigma level.
In contrast, for other parameters, the dark energy effect is almost negligible.
This is because, in the case of larger $f$, density perturbations have not grown maximally by the present time, while for the smaller $f$ case, the perturbations can grow maximally, but their correlation lengths are so small that they are dominated by the standard contribution.

Note that the scale $f\sim 10^{-3.25} \Mpl$, where the effect is the strongest, depends on parameters fixed in our analyses. 
For example, the observational constraints from CMB temperature anisotropies depend on the value of the scale factor at the transition epoch, $a_\ut$, and therefore its preferred value may vary with updates in the observational data. 
Overall, the observational limits on resonance-induced perturbations from the LISW effect depend on both the transition redshift and the decay constant.

We then performed a likelihood analysis using these results. Ideally, an MCMC analysis should be carried out within our model. However, it is unrealistic to run a lattice simulation at each \ac{MCMC} step.
We hence solely compare the likelihood with $\Lambda$CDM for several values of the decay constant $f$.

We used the public Python package \texttt{clipy}~\cite{Planck:2019nip}\footnote{\url{https://github.com/benabed/clipy}} to calculate the likelihood of temperature anisotropies. For $2\leq \ell<30$, we used \texttt{Commander} and for $30 \leq \ell < 2509$, we used \texttt{Plik} as Planck likelihood functions. 

In our setup, $C_\ell^{\mathrm{BG}}$ can be approximated as that in the TDE-CDM universe~\eqref{eq: TDE_EoS} because the transition of \ac{EoS} is sufficiently rapid. We use Planck + DESI + DESY5 MCMC best-fit parameter set $H_0 = 66.33\, \mathrm{km/s/Mpc}$, $\Omega_\mathrm bh^2 = 0.02262$, $\Omega_\mathrm ch^2 = 0.11733$, $A_\mathrm s = 2.129\times 10^{-9}$, $n_\mathrm s = 0.9728$, $\tau = 0.0638$, $w_0 = -0.32$, $a_\ut = 0.952$ and noise parameters.\footnote{
The value of $w_0$ shown here differs significantly from lattice simulations. However, because $w_0$ has a negligible effect on the CMB angular power spectrum, we choose the MCMC best-fit value for $w_0$. We have confirmed that even when $w_0$ is set to $1/6$, the value of the likelihood remains almost unchanged.
}

Log-likelihood for $C_\ell^{\mathrm{BG}}$ is $\ln\mathcal L (C_\ell^{\mathrm{BG}}) = -393.38$. 
This value can also be seen in Sec.~\ref{sec: MCMC analysis}.
We then calculate the likelihood $\ln\mathcal L (C_\ell^{\mathrm{tot}})$ including dark energy perturbations. The differences in the chi-squared values, $\Delta \chi^2 = -2\ln\mathcal L (C_\ell^{\mathrm{tot}}) + 2\ln\mathcal L (C_\ell^{\mathrm{BG}})$, read
\begin{align}\label{eq: chi2}
    \Delta \chi^2 \sim
    \begin{cases}
        0 \quad &(f = 10^{-3}\Mpl), \\
        23.87 \quad &(f = 10^{-3.25}\Mpl), \\
        0.69 \quad &(f = 10^{-3.5}\Mpl), \\
        0.12 \quad &(f = 10^{-4}\Mpl). 
    \end{cases}
\end{align}
As one can see from Table~\ref{tab:MCMC_TDE}, for some parameters, the increase in $\chi^2$ caused by dark energy perturbations exceeds the improvement in $\chi^2$ from $\Lambda$CDM achieved by introducing the TDE background.

In the dark energy model discussed, the LISW signal appears in two ways. First, dark energy perturbations cause an enhancement around multipoles $\ell \sim 50$ (Fig.~11), which could be tested independently using the blurring Sunyaev-Zel'dovich (bSZ) tomography methods in the future \cite{2010A&A...509A..82H}. This method reconstructs the CMB temperature field in redshift shells around galaxy clusters, sampling the anisotropy field across cosmic epochs and helping reduce cosmic variance in single-epoch CMB measurements \cite{2022PhRvD.105f3507I}. However, detecting a signal at $\ell \sim 50$ requires a large, uniform sample of clusters, which demands large sky coverage and precise control of kinetic Sunyaev-Zel'dovich (kSZ) residuals.
Second, our model forecasts a change in the background expansion history at low redshift, $z \sim 0.1$, affecting the late-time evolution of gravitational potentials. This primarily impacts larger angular scales, making it more suitable for bSZ tomography. Low multipoles require fewer clusters for reliable sampling, and the abundance of nearby clusters, along with improved peculiar-velocity reconstruction at low redshift, enhances the feasibility for observational testing.

\subsubsection{Gravitational-wave observations}

\begin{figure*}
	\centering
	\includegraphics[width=0.95\hsize]{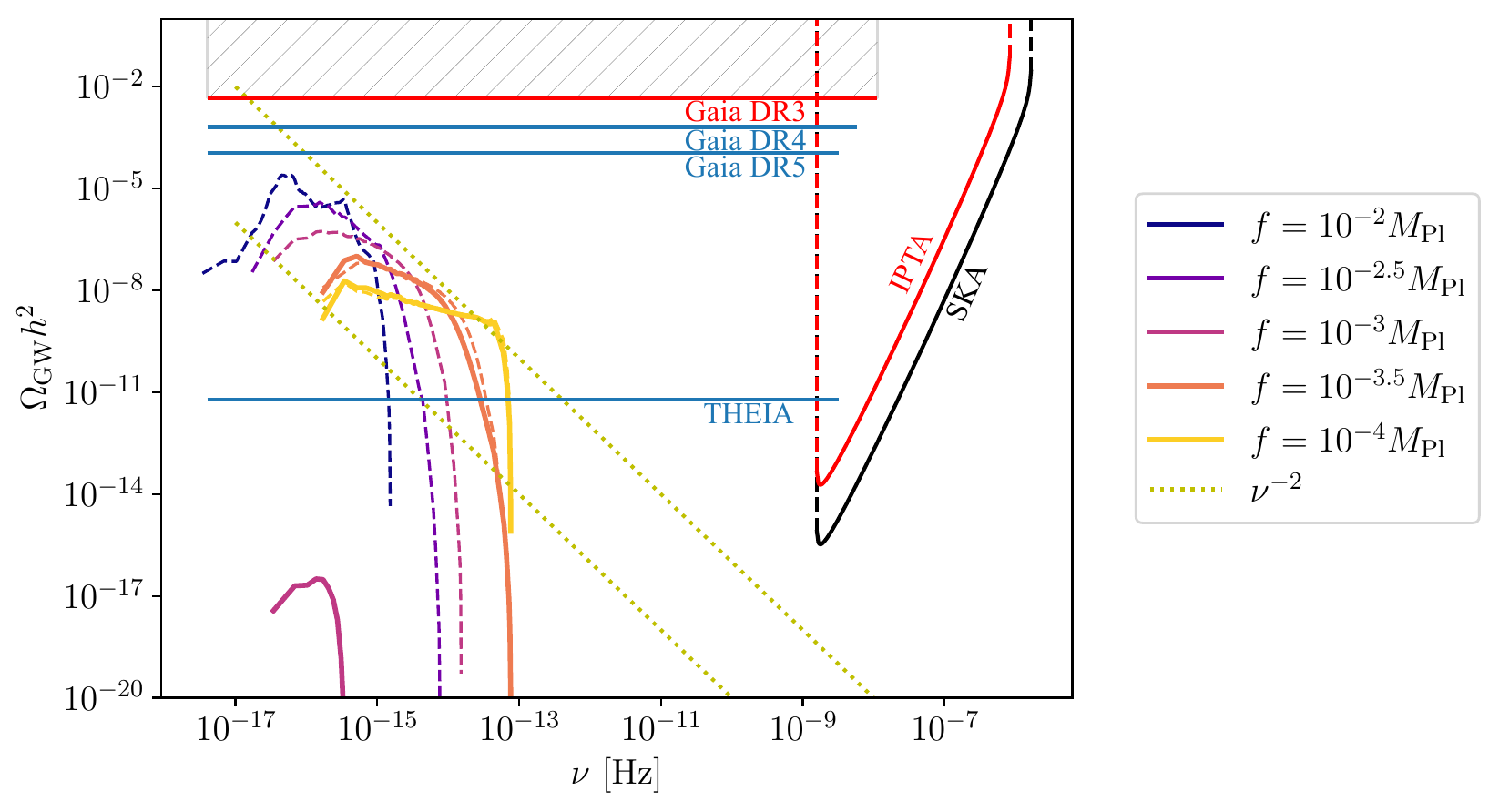}
	\caption{
    Predictions of GW spectra for several values of $f$ from lattice simulations. Thick solid lines represent GWs at the current time $a = 1 $, whereas thin dashed lines represent GWs at sufficiently late times $ a > 1 $ where the \ac{GW} growth has been completed.
    Yellow dotted lines show the $\nu^{-2}$ scaling~\eqref{eq: GW_extrapola} for extrapolation to higher frequencies.  We also show the existing upper bound from quasar astrometry (red horizontal line)~\cite{Darling:2024myz}, future quasar astrometry (Eqs.~\eqref{eq: Gaia_DR4}, \eqref{eq: Gaia_DR5}, and \eqref{eq: THEIA}; blue horizontal lines) and future sensitivity curves~\cite{Schmitz:2020syl} of \ac{PTA}: \ac{SKA} in black and \ac{IPTA} in red. 
	}
	\label{fig: GW}
\end{figure*}

Because the $\mathcal{O}(1)$ fraction of the total energy density becomes the inhomogeneous component at late time around $z \sim 0.1$, GWs are produced substantially, and their redshift from the production to the present is small. In addition, the characteristic frequency scale can be parametrically enhanced from the Hubble scale by a power of $\Mpl / f$, which is advantageous for GW observations. 

Our prediction of the GW spectra is shown in Fig.~\ref{fig: GW}, where existing as well as prospective constraints are plotted. Note that, for too large decay constant $f \gtrsim 10^{-3} \Mpl$, the GW spectrum is still in the process of growth today. This explains why some of the solid lines, representing the present GW strength, do not reach the dashed lines, representing the maximal growth. 

On the other hand, we see a scaling law for the maximally grown GW spectrum (dashed lines) at least in the range $10^{-4} \leq f/\Mpl \leq 10^{-2}$.  Since the required time resolution in the calculation is finer for a lower $f$, the computational cost is higher.  Therefore, we cannot directly simulate the cases of lower $f$.  Instead, we extrapolate our results into smaller values of $f$ as follows. 

Scaling of the GW spectra under the change of the decay constant $f$ can be understood by a dimensional analysis as follows. To this end, let us focus on the characteristic frequency $\nu_\text{peak}$ corresponding to the primary resonance band.  From the equation of motion of the tensor mode $h_{ij}$, we have $\nu_\text{peak}^2 h(\nu_\text{peak}) \sim \frac{\rho}{M_\text{Pl}^2} \sim H_0^2$ because, just after the resonance process, the inhomogeneous component (containing anisotropic stress) of the energy density is comparable to the background energy density of dark energy $\phi$. Then, the GW fraction can be estimated as 
\begin{align}
    \Omega_\text{GW}(\nu_\text{peak}) \sim \frac{\nu_\text{peak}^2 h(\nu_\text{peak})^2}{H_0^2} \sim \left(\frac{H_0 }{ \nu_\text{peak}}\right)^2. 
    \label{eq: GW_extrapola}
\end{align}
Since the characteristic frequency scales as $\nu_\text{peak} \sim k_\text{peak}\sim H_0 \sqrt{M_\text{Pl} /f}$, the GW spectrum becomes smaller and the frequency becomes higher as $f/M_\text{Pl}$ decreases. This scaling is shown by the dotted yellow lines in Fig.~\ref{fig: GW}.

Strictly speaking, the shape of the GW spectrum changes as we change $f$. For smaller $f$, higher resonance bands become more relevant, so the actual peak of the power spectrum of the density perturbations and GWs change from $k_\text{peak}$ and $\nu_\text{peak}$.  This effect was studied, though in the context of preheating after inflation, in Ref.~\cite{Tomberg:2021bll}. They found that the actual peak roughly scales as $k \propto m_\text{th}$ rather than $k\propto m_\text{th} \sqrt{f/\Mpl}$ for $f/\Mpl \lesssim 10^{-4}$. Using $m_\text{th} = \sqrt{V_0}/f$, we see that the actual peak scales as $k \propto H_0 (\Mpl / f)$ for $f/\Mpl \lesssim 10^{-4}$.\footnote{
In Ref.~\cite{Tomberg:2021bll}, $m_\mathrm{th}$ is fixed to fit the CMB normalisation in the context of preheating after inflation, while we fix $V_0$ to fit the present Hubble scale. This explains the apparently different scalings.
}

It is also instructive to consider how far we can extrapolate our result in principle. As we will see shortly, this is related to the UV cutoff of the model.  If we arbitrarily introduced a singular local feature in the potential as in Ref.~\cite{Notari:2024rti} to induce a sudden transitional dynamics, the cutoff scale of the model would be significantly lowered because of unsuppressed coefficients of higher-dimensional operators.  However, the cutoff scale in our model can be hierarchically larger than the relevant physical energy scale. 

The tree-level perturbative unitarity bound~\cite{Lee:1977eg, Logan:2022uus, Bezrukov:2010jz} from scattering of dark energy quanta is 
\begin{align}
    \Lambda  \approx f,
\end{align}
where $\Lambda$ is the UV cutoff scale and we neglected $\mathcal{O}(1)$ numerical factors. On the other hand, the typical energy scale in our physical application is $k_\text{peak}/a = \mathcal{O}(\sqrt{V_0}/\sqrt{f \Mpl}) = \mathcal{O}(H_0 \sqrt{\Mpl / f})$. 
When $f$ is sufficiently small, higher resonance bands become more relevant, and the actual peak energy scales like $\mathcal{O}(H_0 (\Mpl/f))$ as we discussed above. This scale should be below the cutoff scale.  Equating the two scales, we obtain a theoretical lower bound on $f$, 
\begin{align}
    f \gtrsim f_\text{min}  = \sqrt{H_0 \Mpl}  
    \approx 10^{-3} \, \mathrm{eV} \approx 10^{-30} \Mpl. \label{f_min}
\end{align}
The maximum mass scale of $\phi$ at the bottom of the potential coincides with this scale. Note that the scaling $k_\text{peak} \propto m_\text{th} \propto f^{-1}$ has not been tested in the parameter region with such a large extrapolation. In this sense, the uncertainty of $f_\text{min}$ is large. 

A similar lower bound on $f$ can be obtained as follows. As discussed in Refs.~\cite{Lozanov:2019ylm, Tomberg:2021bll}, high frequency (Bunch--Davies) perturbations whose initial energy densities are higher than the background one are not suitable for a semiclassical lattice simulation.
This viewpoint puts an upper bound $k_\mathrm{max}$ on the perturbation wavenumber with 
$(k_\text{max}/a)^4 \lesssim \rho$~\cite{Lozanov:2019ylm}, which leads to $k_\text{max} = \mathcal{O}(\sqrt{H_0\Mpl})$. The peak frequency or wavenumber of GWs scales as $H_0 \Mpl / f$, and it should be less than $k_\text{max}$.  Therefore, the minimum value of $f$ is $H_0 \Mpl / k_\text{max} \sim \sqrt{H_0 \Mpl}$, which reproduces the above estimate~\eqref{f_min}. 

On the other hand, if $f$ is super-Planckian, the transition is not faster than the Hubble time scale, so it cannot mimic the TDE. Therefore, we consider only sub-Planckian $f$.  Super-Planckian $f$ is also challenged by various swampland conjectures (see, \eg, discussions in Ref.~\cite{Tada:2024znt} and references therein).

Let us compare the GW spectra with current and future observational constraints. 
In the frequency range, $\nu \lesssim 10^{-9}\, \mathrm{Hz}$, there is an upper bound on $\Omega_\text{GW}$ from quasar astrometry~\cite{Jaraba:2023djs, Darling:2024myz} since the presence of the stochastic GW background affects the angular positions of quasars in a correlated way. In this low-frequency range, it is impossible to see the oscillation signals because the oscillation period is much larger than the practical observational time, and hence it cannot be used for the detection of \acp{GW}.  
Only an upper bound can be placed due to the absence of a linear shift in the angular position of quasars.  The latest and strongest bound reads~\cite{Darling:2024myz}
\begin{align}
    \Omega_\text{GW}h^2 \leq 4.7 \times 10^{-3} \quad (\text{Gaia DR3 (34\,mo})), 
    \label{eq: Gaia_DR3}
\end{align}
for $4.0 \times 10^{-18} \, \mathrm{Hz} \leq \nu \leq 1.12 \times 10^{-8} \, \mathrm{Hz}$,\footnote{
Following Refs.~\cite{Darling:2018hmc, Jaraba:2023djs}, we conservatively adopt the lower cutoff of the frequency to correspond to the $25$th percentile of the distance distribution of the quasars rather than the $75$th percentile adopted in Ref.~\cite{Darling:2024myz}. 
} where $h = H_0/(100\, \mathrm{km}/\mathrm{s}/\mathrm{Mpc})$ is the reduced Hubble constant. For a higher frequency, the signal is cancelled out by fast oscillations in the current data, where time-series information is not available, while for a lower frequency, the distant-source limit used in the analyses is no longer valid.   The above constraint is shown by the horizontal red line and gray shading in Fig.~\ref{fig: GW}.  The constraint is based on Gaia~\cite{2016A&A...595A...1G} DR3~\cite{2023A&A...674A...1G}, which amounts to the data for 34 months.  

Let us follow Refs.~\cite{Garcia-Bellido:2021zgu, Jaraba:2023djs} to estimate the improvement of the bound in future. With a longer observation period $T$, new quasars would be discovered. Even if we conservatively fix the number of quasars, longer observational time allows us to put a tighter constraint on $\Omega_\text{GW}$ proportional to $T^{-3}$~\cite{Jaraba:2023djs}. With this conservative improvement, we obtain prospective constraints
\begin{align}
    \Omega_\text{GW}h^2 &\leq  6.5 \times 10^{-4} \quad (\text{Gaia DR4 (5.5\,yr})), 
    \label{eq: Gaia_DR4} \\ 
    \Omega_\text{GW}h^2 &\leq  1.1 \times 10^{-4} \quad (\text{Gaia DR5 (10\,yr})).  
    \label{eq: Gaia_DR5}
\end{align}
Finally, assuming 60-fold improvement of angular resolution and 100-fold increase of the number of quasars (with low uncertainty of proper motion~\cite{Darling:2024myz})~\cite{Garcia-Bellido:2021zgu, Jaraba:2023djs} by 
Telescope for Habitable Exoplanets and Interstellar/Intergalactic Astronomy (THEIA)~\cite{2017arXiv170701348T, 10.3389/fspas.2018.00011}, the angular correlation method in Ref.~\cite{Darling:2024myz} leads to a $60^2 \times \binom{100}{2}$ times tighter prospective constraint, where $\binom{100}{2}$ is a binomial coefficient, namely,
\begin{align}
    \Omega_\text{GW}h^2 \leq 6.0 \times 10^{-12} \quad (\text{THEIA (optimistic)}).
    \label{eq: THEIA}
\end{align}
With such a futuristic constraint, some portion of the parameter space in our model can be excluded. 
Note that this type of observation is not sensitive to the frequency of the GWs since it is degenerate with the amplitude. In this sense as well, this observational channel is not particularly suitable for the discovery of the GW signals. 

For a higher frequency range, $\nu \gtrsim 10^{-8}\,\mathrm{Hz}$, the astrometry method can also put a tight constraint on $\Omega_\text{GW}$ (once the time series data are released). For prospective constraints by Gaia and Theia, see Ref.~\cite{Garcia-Bellido:2021zgu}.

There are also constraints from \acfp{PTA}. Our signals are many orders of magnitude below the existing constraints~\cite{EPTA:2015qep, NANOGRAV:2018hou, Shannon:2015ect} and reported signals~\cite{NANOGrav:2023gor, EPTA:2023fyk, Reardon:2023gzh, Xu:2023wog, InternationalPulsarTimingArray:2023mzf}.  We show the prospective constraints from the \acf{IPTA}~\cite{2010CQGra..27h4013H, 2013CQGra..30v4010M, Perera:2019sca} and the \acf{SKA}~\cite{Carilli:2004nx, Janssen:2014dka, Weltman:2018zrl} in red and black, respectively, in Fig.~\ref{fig: GW}. Note, however, that the realistic sensitivity is weaker than these because the reported signals around the nanohertz range serve as the background to our signals (see Ref.~\cite{Babak:2024yhu}).
Comparing the PTA sensitivity curves and the yellow dotted lines representing the extrapolation of the signal for smaller $f$, we see that it will be difficult to observe the GWs by PTAs in our scenario.

Here, we discussed the astrometric and PTA constraints on the GWs produced by the resonance of the quintessence field. 
The astrometric and PTA observations can also directly probe a light scalar field such as the quintessence. This and various other observational implications are discussed below.

\subsubsection{Discussion on other observational implications}\label{sec:other_obs}

There are some other possibilities to observe the resonant phenomenon or its implications in our scenario.

\paragraph{Gravitational effects of dark energy oscillations.} After the dark energy field $\phi$ becomes an additional component of dark matter, it oscillates on a cosmological time scale with its frequency proportional to $m_\phi$.  Though significant inhomogeneity is generated, there is a residual homogeneous component. Gravitational effects of such oscillations are searched for observationally.  PTAs put constraints on the ultralight scalar-field dark matter~\cite{Khmelnitsky:2013lxt, EPTA:2023xxk, NANOGrav:2023hvm}.  The ultralight scalar-field being 100\% dark matter is excluded around $10^{-24} \, \mathrm{eV} \lesssim m_\phi \lesssim 10^{-23}\, \mathrm{eV}$~\cite{EPTA:2023xxk}.\footnote{
It is discussed that a heavier mass range can also be constrained~\cite{Kim:2023kyy}, but the assumptions of virialization should be revisited to constrain our scenario. See also Ref.~\cite{Kim:2023pkx} and Ref.~\cite{Kim:2024xcr} for the application of similar ideas to interferometers and astrometry, respectively.}  
Thus, the direct PTA constraint on an ultralight scalar field can be stronger than the PTA constraint on GWs that is produced by the dynamics of the scalar field for this mass range.  Effects of inhomogeneous components need to be studied to constrain our scenario.

\paragraph{Gravitational effects of dark energy inhomogeneity.} After the end of the growth of perturbations, about half of the dark energy density behaves as an additional dark matter component. This second species of dark matter is highly inhomogeneous and may have some observational implications. As we mentioned earlier, oscillons may form in our setup.  According to Ref.~\cite{Lozanov:2017hjm} in the context of preheating after primordial inflation, oscillons form with the same ($\tanh^2$) potential. In our case, there is a residual matter component ($\chi$), so the fate of the dark energy perturbations may be different from that in their setup. Compact objects such as oscillons can be probed by the gravitational lensing effect, depending on the size and the number density of such compact objects.  

Whether or not oscillons are formed, significant inhomogeneity of $\phi$ generates the gravitational potential, which attracts dark as well as visible matter.  This affects the velocity distribution of dark matter and the peculiar velocities of galaxies.  This in turn can affect the \ac{CMB} via the kSZ effect.  Taking the correlation between the CMB and the galaxy density, one can tomographically estimate the radial velocity field (kSZ tomography~\cite{2011MNRAS.413..628S}). Using this effect, one can study the effect of dark energy perturbations~\cite{Adolff:2025fli}. In contrast to the case studied in Ref.~\cite{Adolff:2025fli}, the sound horizon of the dark energy field shrinks in our scenario, and significant inhomogeneity of dark energy is produced, both of which enhance the subhorizon velocity perturbations. The effects of transient dark energy in our scenario may be probed for sufficiently small $f$ that induces a velocity field on sufficiently large scales that can be probed by observations.  It will be interesting to study these effects.

\paragraph{GW-induced curl mode in the weak lensing.} GWs are produced at late times, such as around $z \simeq 0.1$ in our scenario, so it does not affect the last scattering of the CMB. This means that the usual $B$-mode polarization constraint on the tensor-to-scalar ratio does not apply to our scenario. However, GWs produced after recombination give rise to the rotation or curl mode of weak gravitational lensing. This affects the galactic images and the CMB~\cite{Sarkar:2008ii, Namikawa:2014lla}. Since the effects are more significant on large scales, it is sensitive to the low-frequency GWs.  In particular, precise CMB measurements can probe GWs produced at late time, $z < 10$, with a low frequency, $k \lesssim 10^{-3} \,\mathrm{Mpc}^{-1}$~\cite{Namikawa:2014lla},  which corresponds to large $f = \mathcal{O}(\Mpl)$ in our model. Unfortunately, GWs have not grown significantly for $f > 10^{-3} \Mpl$ and the transition redshift $z_\text{t} \lesssim 0.1$.  

\paragraph{Violation of the consistency relations for correlation functions.}
There are consistency relations among primordial cosmological correlation functions~\cite{Maldacena:2002vr, Creminelli:2004yq} assuming single-field dynamics during inflation.  In our scenario, additional density perturbations and \acp{GW} are produced at late times, independently of CDM density perturbations, which can lead to violations of the consistency relations. For a test of the consistency relations in the Large Scale Structure of the Universe~\cite{Peloso:2013zw, Kehagias:2013yd, Creminelli:2013mca}, see Ref.~\cite{Sugiyama:2023zvd}. 

\paragraph{Effects on black holes}
Around a rotating black hole, an ultralight scalar field in a resonance mass range is enhanced by superradiance, which occurs independently of the cosmological abundance of the scalar.  This process extracts the angular momentum from the Kerr black hole significantly unless the scalar self-interaction is too strong to impede the resonant growth.  Observations of existing rotating black holes place constraints on the possible mass range of an ultralight scalar field. The mass ranges around $ \mathcal{O}(10^{-19})\,\mathrm{eV}$ and $\mathcal{O}(10^{-13})\,\mathrm{eV}$ to $\mathcal{O}(10^{-12})\, \mathrm{eV}$ are constrained~\cite{Baryakhtar:2020gao, Unal:2020jiy, Hoof:2024quk, Witte:2024drg} (see also Ref.~\cite{Palomba:2019vxe}). 

If a cloud of an ultralight scalar field exists around merging binary black holes, the \ac{GW} signal can be modified (see, e.g., Refs.~\cite{Baumann:2022pkl, Boudon:2023vzl, Aurrekoetxea:2024cqd}).  In our scenario, clumps of the quintessence are formed by the tachyonic resonance at around $z \sim 0.1$, so only the binary mergers after that time can be affected. It would be interesting if such a redshift-dependent modification of the GW signals is found. 

\paragraph{Cosmic birefringence and astrophysical constraints in extended models.}
While a natural interpretation of $\phi$ with an exponential-type potential is a dilaton-like field, an axion-like field can have a similar flat potential in the (dark) pure Yang--Mills theory with the Chern--Simons coupling~\cite{Nomura:2017ehb}. Our potential $V(\phi)$ may have such an origin.  It is also interesting to note that the Chern--Simons coupling to photons can explain~\cite{Fujita:2020ecn, Berghaus:2020ekh, Fung:2021wbz, Nakagawa:2021nme, Jain:2021shf, Choi:2021aze, Gasparotto:2022uqo} the cosmic birefringence~\cite{Minami:2020odp, Diego-Palazuelos:2022dsq, Eskilt:2022wav, Eskilt:2022cff, Cosmoglobe:2023pgf}.\footnote{
One may naively wonder that photons would be copiously produced from such a coupling, and the model would be excluded.  
However,  photons acquire the plasma mass, and their production from the scalar field $\phi$ is suppressed when $m_\phi$ is below the plasma frequency~\cite{Alonso-Alvarez:2019ssa}.
} 
Simultaneous explanation of cosmic birefringence and dynamical dark energy was discussed in Ref.~\cite{Tada:2024znt}. 
Therefore, it is also natural to identify $\phi$ as an axion-like field and consider the Chern--Simons coupling to some gauge fields.

Let us first consider the photon coupling, which is motivated by the cosmic birefringence.  
Although the energy density of $\phi$ develops its inhomogeneity, $\langle \phi \rangle$ averages out in the oscillation regime. Therefore, assuming a sufficiently homogeneous initial condition, the constraint from anisotropic cosmic birefringence~\cite{BICEP2:2017lpa, Namikawa:2020ffr} can be satisfied. Interestingly, there is a proposal of a tomographic method to probe the redshift-dependence of the cosmic birefringence~\cite{Naokawa:2025shr}, which may distinguish our scenario from others. 
There are also other constraints, mainly from astrophysical considerations, on the photon coupling as well as on various couplings to the Standard-Model particles (see, \eg, Refs.~\cite{AxionLimits, Caputo:2024oqc}).

It is also possible that $\phi$ couples to dark gauge bosons such as dark photons. 
The field variation of $\phi$ induces the tachyonic instability of the dark photon depending on its mass, so the dark photon can be substantially produced.  They can constitute dark matter or dark radiation. 
If the dark photon mixes with the Standard Model photon, the effective coupling constants are again constrained by astrophysical considerations.  

\paragraph{Cosmological tensions.}
Since the cosmology is significantly altered at very late time at $z = \mathcal{O}(0.1)$ in our scenario, it would be natural to discuss possible implications on the cosmological tensions such as the Hubble tension~\cite{Verde:2019ivm, H0DN:2025lyy, DiValentino:2021izs, Perivolaropoulos:2021jda, Abdalla:2022yfr} and the $S_8$ tension~\cite{Perivolaropoulos:2021jda, Abdalla:2022yfr}. See, \eg, Refs.~\cite{Seto:2024cgo, Jiang:2024xnu, RoyChoudhury:2024wri, Colgain:2024mtg, Toda:2025dzd} discussing the cosmological tensions and the recent DESI results. 
The Hubble tension becomes severer for quintessence models~\cite{Banerjee:2020xcn, Lee:2022cyh}.  In our scenario, there are significant inhomogeneities produced at late time for the field $\phi$, which initially behaves like dark energy and later behaves like dark matter, so the situation is more complicated. 
The inhomogeneities are a candidate ingredient to potentially ameliorate or solve the tensions. 
While the comprehensive discussions on the implications on the cosmological tensions are beyond the scope of our work, it will be interesting to study them elsewhere. 

\section{Conclusion}\label{sec:conclusion}

In this paper, we have studied a quintessence model that realises dynamical dark energy with a sudden transitional feature (the TDE in a broad sense). Despite the sharp transitional feature, our potential is smooth and regular.  This is an alternative to the usual quintessence models that can be approximated by the DSCh approach.  As we have discussed in Sec.~\ref{sec:toy_model} (and Appendix~\ref{sec: FUll MCMC} in more detail), the quality of the fit to the observational data in the TDE model is comparable to that in the DSCh approach. In Sec.~\ref{sec:quintessence}, we have studied the background dynamics as well as the growth of perturbations in detail, both at the analytic and perturbative level and at the numerical and nonlinear level. Notably, the tachyonic resonance inevitably occurs in our model, which not only governs the development of perturbations but also backreacts to the homogeneous background. This leads to an interesting profile of observables such as the \ac{EoS} parameter and the Hubble parameter. 

In addition to the dark energy phenomenology, we have found rich observational implications originating from the significant inhomogeneity produced by the tachyonic resonance. Such inhomogeneities generate large curvature perturbations, which affect the \ac{CMB} through the late ISW effect. We have found that the effect is significant in a limited range in the parameter space near $f \sim 10^{-3.25} \Mpl$. Such a deviation is an interesting target for future \ac{CMB} observations. Another interesting signal is the GWs sourced by the inhomogeneity of the quintessence field. We have seen that future quasar astrometry is a useful tool to constrain this scenario. Other observational implications have been commented on in Sec.~\ref{sec:other_obs}. 


Many qualitative features of our model will be shared by different implementations of our basic ideas.  
In this paper, we have fixed the quintessence model to be a single-field scalar field model with a specific $\tanh^2$ potential for concreteness.  However, our main idea to reproduce the sudden transitional feature in $w(a)$ can be more generally applicable to various models.  For example, one can imagine a multi-field model which has an instability like the waterfall instability in hybrid-inflation models~\cite{Linde:1993cn}. Such models will also exhibit the tachyonic instability and particle production.  
Another possibility of a sudden transition is a phase transition in the dark sector. In this case, tachyonic instability may not be involved.  Instead, it may involve bubble collisions and production of gravitational waves if the transition is of the first order.  

While we have focused on the quintessence model, it is also possible to generalise our scenario to include the (effective) phantom behaviour.  A simple extension in such a direction is to introduce interactions between dark energy and dark matter along the line of Refs.~\cite{Huey:2004qv, Das:2005yj, Khoury:2025txd}. It will be interesting to study how much the fit to the observational data is improved when we have both the phantom regime and a sharp transitional feature. 

One of the future directions is a thorough exploration of the full parameter space $(f, V_0, \phi_\ui)$ in our model, while we have dealt with the first parameter as the main free parameter and fixed the latter two parameters to reproduce the target values of $H_0$ and the transition redshift in this paper. Because of the numerical cost of the lattice calculation, we have not performed the MCMC analyses in our quintessence model. Machine-learning techniques may be useful in this direction.  

In conclusion, the dynamical dark energy model we have studied in this paper has rich observational implications, some of which can be probed in future observations. Interestingly, the cosmology is significantly modified at very late times around $z \sim \mathcal{O}(0.1)$ with possibly further interesting observational consequences to be explored.

\acknowledgments

We thank Shintaro Hayashi and Hironao Miyatake for their collaborations at the early stage of this work.
SY thanks Shintaro Hayashi for helpful advice and support with the modification of \texttt{CAMB} and the compilation of \texttt{cobaya}.
TT thanks the Yukawa Institute for Theoretical Physics at Kyoto University; discussions during the YITP workshop YITP-W-25-10 on ``Progress in Particle Physics 2025'' were useful to complete this work. TT thanks Gabriele Franciolini, Kai Murai, Kimihiro Nomura, and Wen Yin for useful discussions and conversations.  TT was supported by the 34th Academic research grant FY 2024 (Natural Science) No.~9284 from DAIKO FOUNDATION. 
This work is supported in part by the JSPS grant numbers 21H04467 (KI), 24K00625 (KI), and 24K07047 (YT),
and JST FOREST Program JPMJFR20352935 (KI).

\appendix

\section{Linear analysis of the tachyonic resonance}\label{sec: linear analysis}

In this appendix, we study tachyonic resonance in the linear regime, following Ref.~\cite{Tomberg:2021bll}. In Sec.~\ref{ssec:no_expansion_limit}, we review the Floquet analysis neglecting the Hubble expansion. The Hubble expansion is reintroduced in Sec.~\ref{ssec:with_matter}, taking into account the existence of matter components. 

\subsection{\boldmath $H\to0$ limit}\label{ssec:no_expansion_limit}

Let us first neglect the cosmic expansion for simplicity.
The background \ac{EoM} for the homogenous mode $\phi=\phi_0(t)$ is given by
\bae{
  \ddot{\phi}_0+V'(\phi_0)=0,
}
where the dot denotes the derivative with respect to the cosmic time $t$.
Defining a characteristic mass scale $m_\uth$ via $V_0=m_\uth^2f^2$, we renormalise variables as
\bae{
    \varphi=\frac{\phi_0}{f} \qc
    \calV(\varphi)=\frac{V(\phi_0)}{m_\uth^2f^2}=\tanh^2\varphi \qc \tau=m_\uth t,
}
and the background equation
\bae{
    \partial_\tau^2\varphi(\tau)+\calV'\qty(\varphi(\tau))=0,
}
where a prime denotes differentiation with respect to the argument. 
For the initial condition $\varphi(0)=\varphi_\amp$ and $\varphi'(0)=0$, the solution is given by
\bae{
    \varphi(\tau)= \arcsinh\pqty{\abs{\sinh\varphi_\amp}\cos\frac{\pi\tau}{\Delta\tau}},
}
where $\Delta \tau$ is the half period of an oscillation of $\varphi$, i.e.
\bae{
    \Delta\tau=\frac{\pi\cosh\varphi_\amp}{\sqrt{2}}.
}

Along this background solution, the (dimensionless) linear \ac{EoM} for the Fourier mode of the perturbation, $\delta\varphi_\kappa$, is given by
\bae{
    \partial_\tau^2\delta\varphi_\kappa(\tau)+\bqty{\kappa^2+\calV''\qty(\varphi(\tau))}\delta\varphi_\kappa(\tau)=0, \label{eq: linear EoM}
}
where the dimensionless wavenumber $\kappa=k/m_\uth$ is related to the ordinary wavenumber $k$.
As the background oscillates, $\calV''(\varphi(\tau))$ exhibits periodicity in time and drives the resonance of the perturbation.
Its periodic average and variance are given by
\bae{
    \expval{\calV''}=\sech\varphi_\amp\pqty{3\sech^2\varphi_\amp-1}, 
}
and 
\bae{
    \sigma_{\calV''}^2&=\expval{(\calV'')^2}-\expval{\calV''}^2 \nonumber \\
    &=\bmte{\frac{\sech^7\varphi_\amp\sinh^4(\varphi_\amp/2)}{8} \\
    \times\left(447+468\cosh\varphi_\amp +92\cosh(2\varphi_\amp) \right. \\
    \left. +12\cosh(3\varphi_\amp) +5\cosh(4\varphi_\amp)\right).}
}
The bracket represents the periodic average defined by
\bae{
    \expval{\calO(\tau)}=\frac{1}{\Delta\tau}\int_0^{\Delta\tau}\calO(\tau)\dd{\tau}.
}
Note that the period of $\calV''$ is $\Delta\tau$, the half period of $\varphi(\tau)$, due to the $\mathbb{Z}_2$ symmetry of the potential.

\begin{figure}
	\centering
	\includegraphics[width=0.95\hsize]{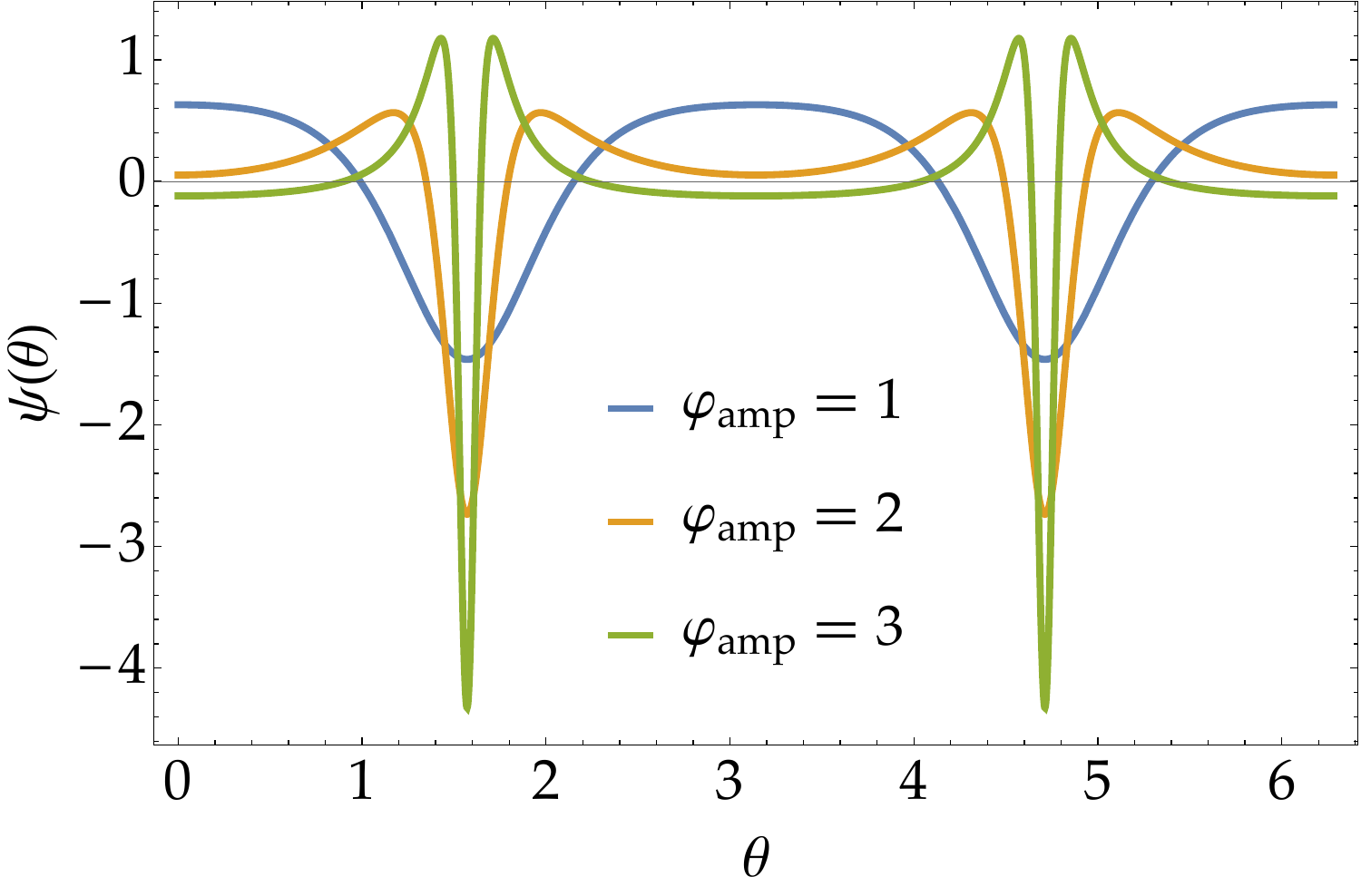}
	\caption{The normalised periodic function $\psi(\theta)$~\eqref{eq: psi} for $\varphi_\amp=1$ (blue), $2$ (orange), and $3$ (green).
	}
	\label{fig: psi}
\end{figure}

Following the prescription of Ref.~\cite{Fukunaga:2019unq}, we define the normalised periodic function $\psi(\theta)$ by
\bae{\label{eq: psi}
	\psi(\theta)=-\frac{\calV''(\varphi(\theta\Delta\tau/\pi))-\expval{\calV''}}{\sqrt{2}\sigma_{\calV''}},
}
so that
\bae{
    \expval{\psi(\theta)}=0 \qc \expval{\psi^2(\theta)}=\frac{1}{2}.
}
We employ the phase parameter $\theta=\pi\tau/\Delta\tau$ as a renormalised time variable.
Then, the perturbation equation~\eqref{eq: linear EoM}, the second-order differential equation with a periodic mass term, a.k.a the Hill equation, can be viewed as a generalisation of the Mathieu equation as
\bae{\label{eq: Hill}
	\partial_\theta^2\delta\varphi_\kappa\qty(\tau(\theta))+\bqty{A_\kappa-2q\psi(\theta)}\delta\varphi_\kappa\qty(\tau(\theta))=0,
}
where 
\bae{\label{eq: Ak and q}
	A_\kappa=\pqty{\kappa^2+\expval{\calV''}}\pqty{\frac{\Delta\tau}{\pi}}^2 \qc q=\frac{\sigma_{\calV''}}{\sqrt{2}}\pqty{\frac{\Delta\tau}{\pi}}^2.
}
If the periodic function $\psi$ is sinusoidal, \ie, $\psi(\theta)=\cos(2\theta)$, it reduces to the Mathieu equation, whose resonance structure is determined by two parameters $A_\kappa$ and $q$ and well studied in the literature~\cite{mclachlan1964theory}.
In our case, as shown in Fig.~\ref{fig: psi}, the periodic function $\psi(\theta)$ has a non-trivial form depending on $\varphi_\amp$, so the resonance structure should be investigated for each value of $\varphi_\amp$.

\begin{figure}
	\centering
    \begin{tabular}{l}
		\includegraphics[width=0.98\columnwidth]{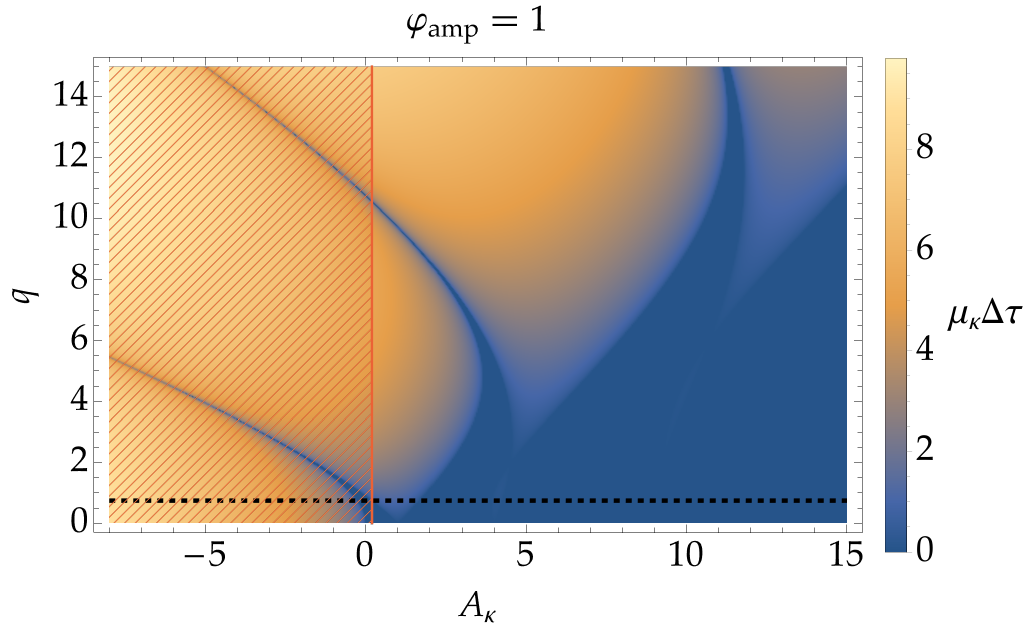}	\\		
		\includegraphics[width=1\columnwidth]{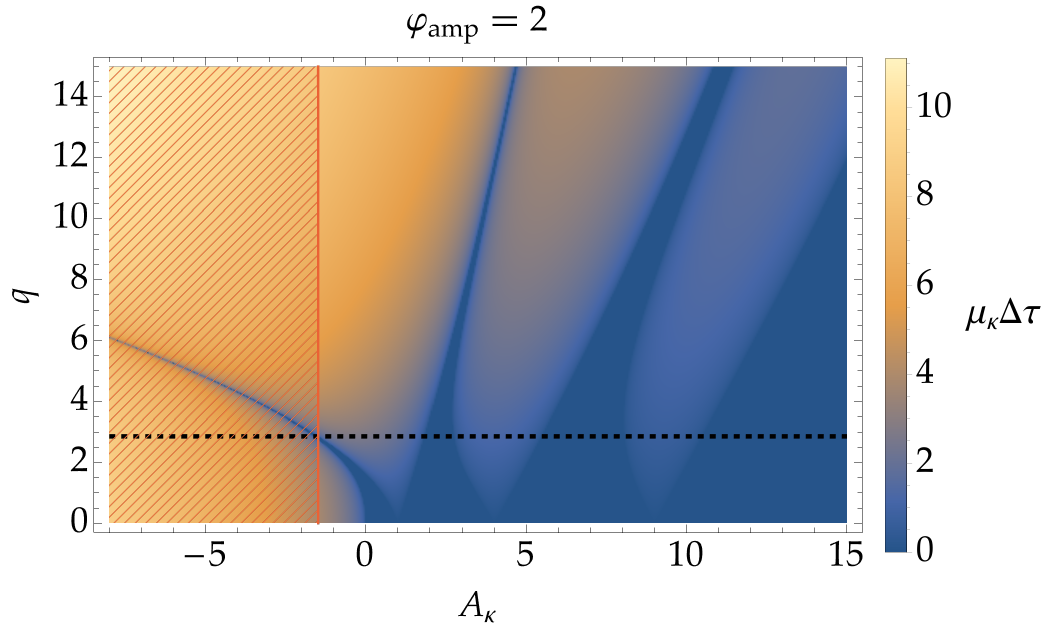}\\
		\includegraphics[width=0.98\columnwidth]{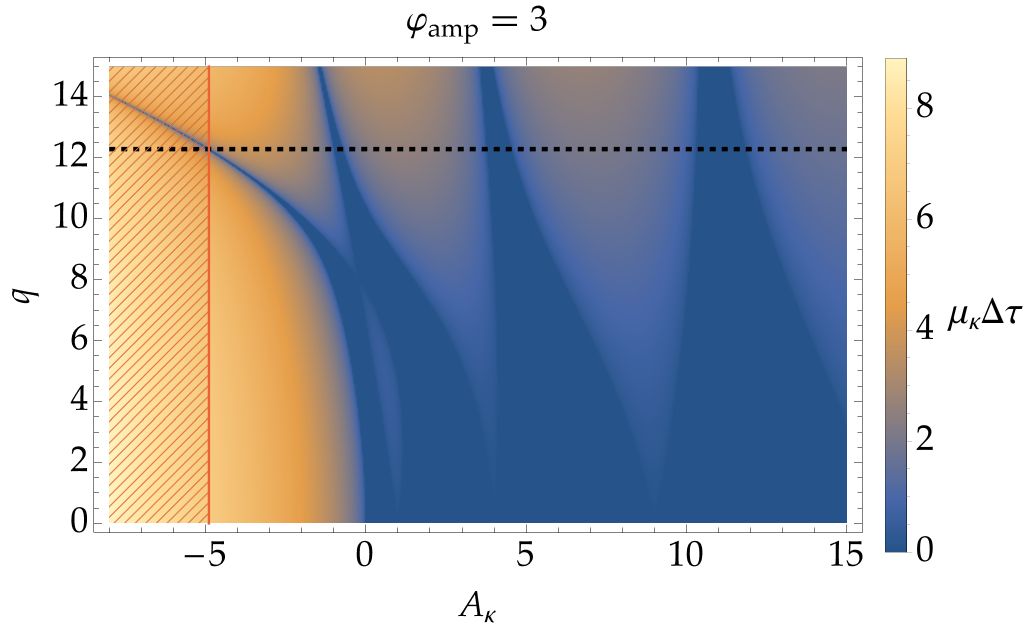}
	\end{tabular}
	\caption{The growth rates $\mu_\kappa\Delta\tau$ as functions of $A_\kappa$ and $q$ for $\varphi_\amp=1$ (top), $2$ (middle), and $3$ (bottom). 
    The horizontal dashed lines are $q=\frac{\sigma_{\calV''}}{\sqrt{2}}\pqty{\frac{\Delta\tau}{\pi}}^2$~\eqref{eq: Ak and q} and the red-hatched regions are forbidden due to $\kappa^2<0$ according to the relation~\eqref{eq: Ak and q} in this model.
	}
	\label{fig: Floquet}
\end{figure}

Along this Hill equation, one seeks periodic solutions up to a growth factor:
\bae{
    \delta\varphi_\kappa(\tau+\Delta\tau)=\ee^{\lambda_\kappa\Delta\tau}\delta\varphi_\kappa(\tau),
}
where $\lambda_\kappa$ is called the Floquet exponent.
Among these solutions, we are interested in the maximum growth rate $\mu_\kappa$ defined by 
\bae{
	\mu_\kappa=\max\Re\lambda_{\kappa}.
}
For each exponent $\lambda_\kappa$, the growth factor $\ee^{\lambda_\kappa\Delta\tau}$ can be calculated as an eigenvalue of the matrix $G$ defined by 
\bae{
	w(\tau+\Delta\tau)=w(\tau)G \qc
	w(\tau)=\pmqty{u_1(\tau) & u_2(\tau) \\
		u_1'(\tau) & u_2'(\tau)},
}
with any two independent solutions $u_1$ and $u_2$ of the Hill equation~\eqref{eq: Hill}.
Practically, the choice $w(0)=I$ is useful so that $G=w(\Delta\tau)$, where $I$ denotes the unit matrix.
Furthermore, the \ac{EoM} ensures the conservation of the Wronskian $\det w$ in time. It implies $\det G=1$ and hence $\lambda_{\kappa+}=-\lambda_{\kappa-}$ for two exponents $\lambda_{\kappa\pm}$.
They can be obtained as
\bae{
  \lambda_{\kappa\pm}\Delta\tau=\pm\arccosh\qty(\frac{1}{2}\tr G).
}
If $\abs{\tr G}<2$, $\lambda_{\kappa\pm}$ are pure imaginary and no enhancement occurs.
If $\abs{\tr G}>2$, $\lambda_{\kappa\pm}$ are real and the perturbation grows exponentially, $\propto\exp(\mu_\kappa\Delta\tau)=\exp(\Re\lambda_{\kappa+}\Delta\tau)$.
In Fig.~\ref{fig: Floquet}, we exemplified $\mu_\kappa\Delta\tau$ as functions of $A_\kappa$ and $q$ for several values of $\varphi_\amp$.

Practically, $A_\kappa$ and $q$ are related to $\varphi_\amp$ via Eq.~\eqref{eq: Ak and q} in this model. These relations are illustrated by dashed lines in Fig.~\ref{fig: Floquet}.
Figure~\ref{fig: Floquet chart} shows the growth rate as a function of $\varphi_\amp$ and $\kappa$ through the relations.

\subsection{With matter components}\label{sec: with matter}\label{ssec:with_matter}

Let us then include the matter components and turn on the cosmic expansion.
The background equations are modified as
\bege{\label{eq: bg EoM w/ matter}
    \ddot{\phi}_0+3H\dot{\phi_0}+V'(\phi_0)=0 \qc \dot{z}=-(1+z)H, \\
    3\Mpl^2H^2=\rho_\um+\frac{\dot{\phi}_0^2}{2}+V(\phi_0) \qc \rho_\um=\rho_{\um0}(1+z)^3.
}
The matter energy density $\rho_\um$ with the current value $\rho_{\um0}$ is merely added as an external field being diluted as $\propto(1+z)^3$ with the redshift $z$.

\begin{figure*}
	\centering
	\begin{tabular}{c}
		\begin{minipage}{0.48\hsize}
			\centering
			\includegraphics[width=\hsize]{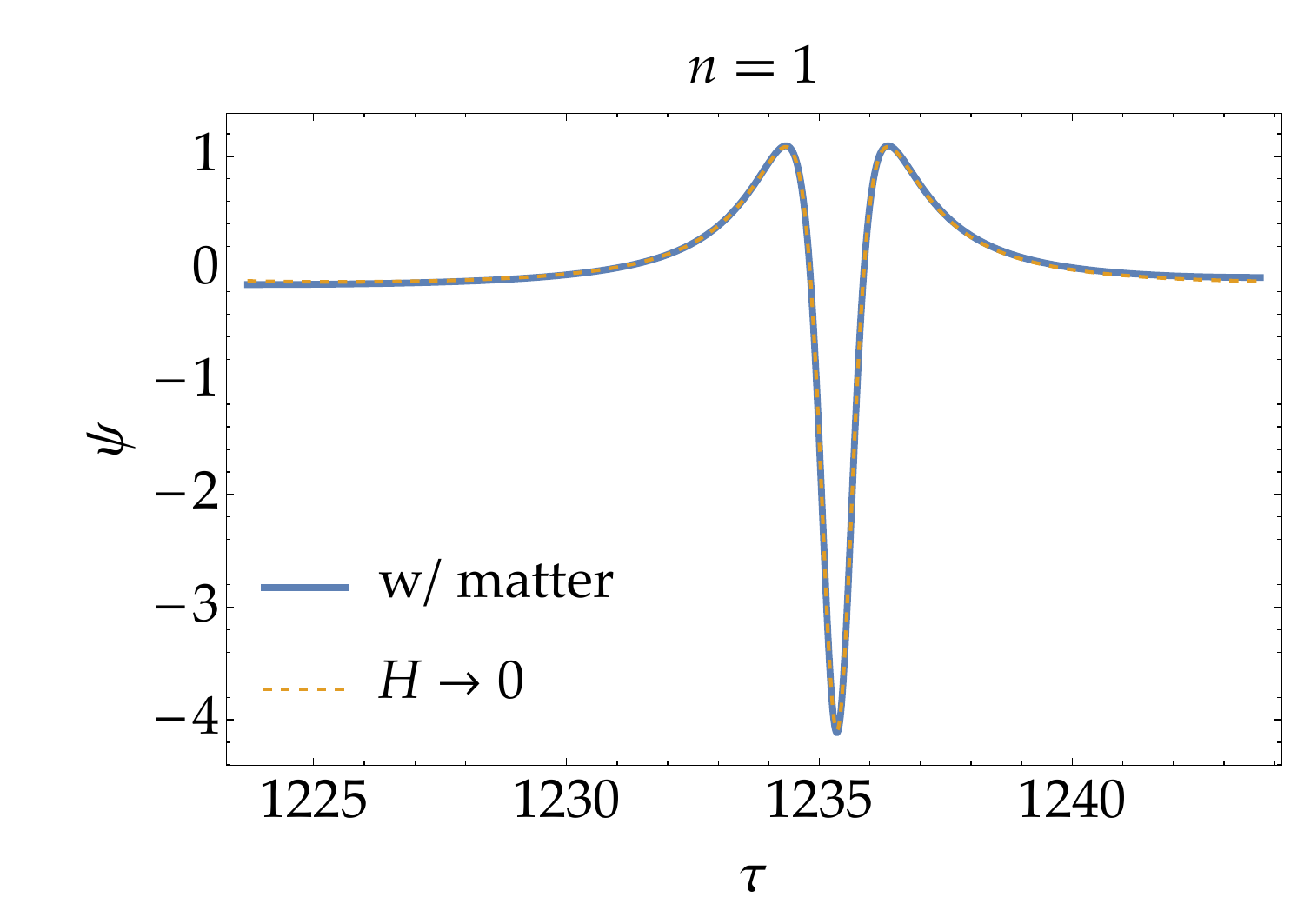}
		\end{minipage}
		\begin{minipage}{0.48\hsize}
			\centering
			\includegraphics[width=\hsize]{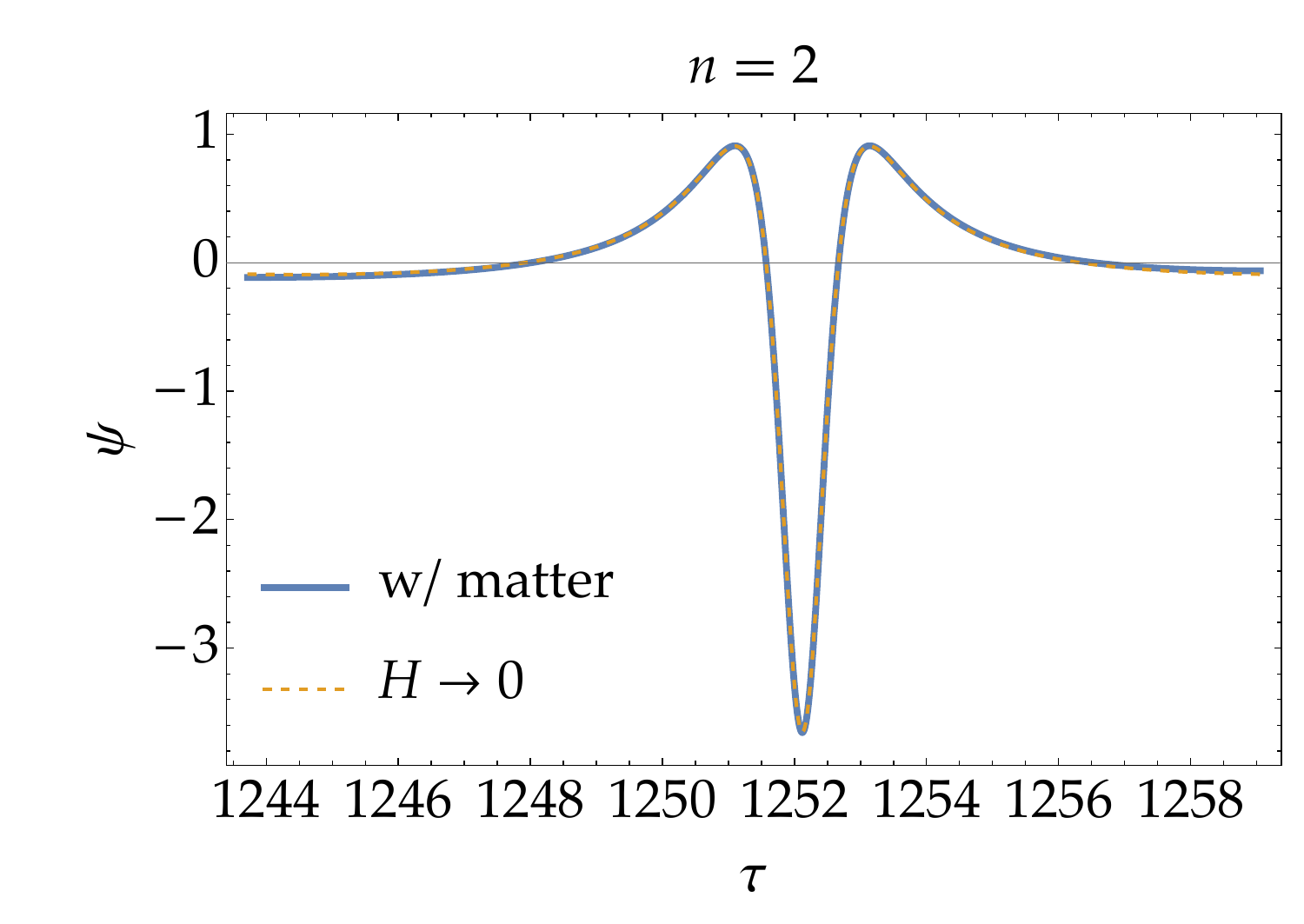}
		\end{minipage} \\
		\begin{minipage}{0.48\hsize}
			\centering
			\includegraphics[width=\hsize]{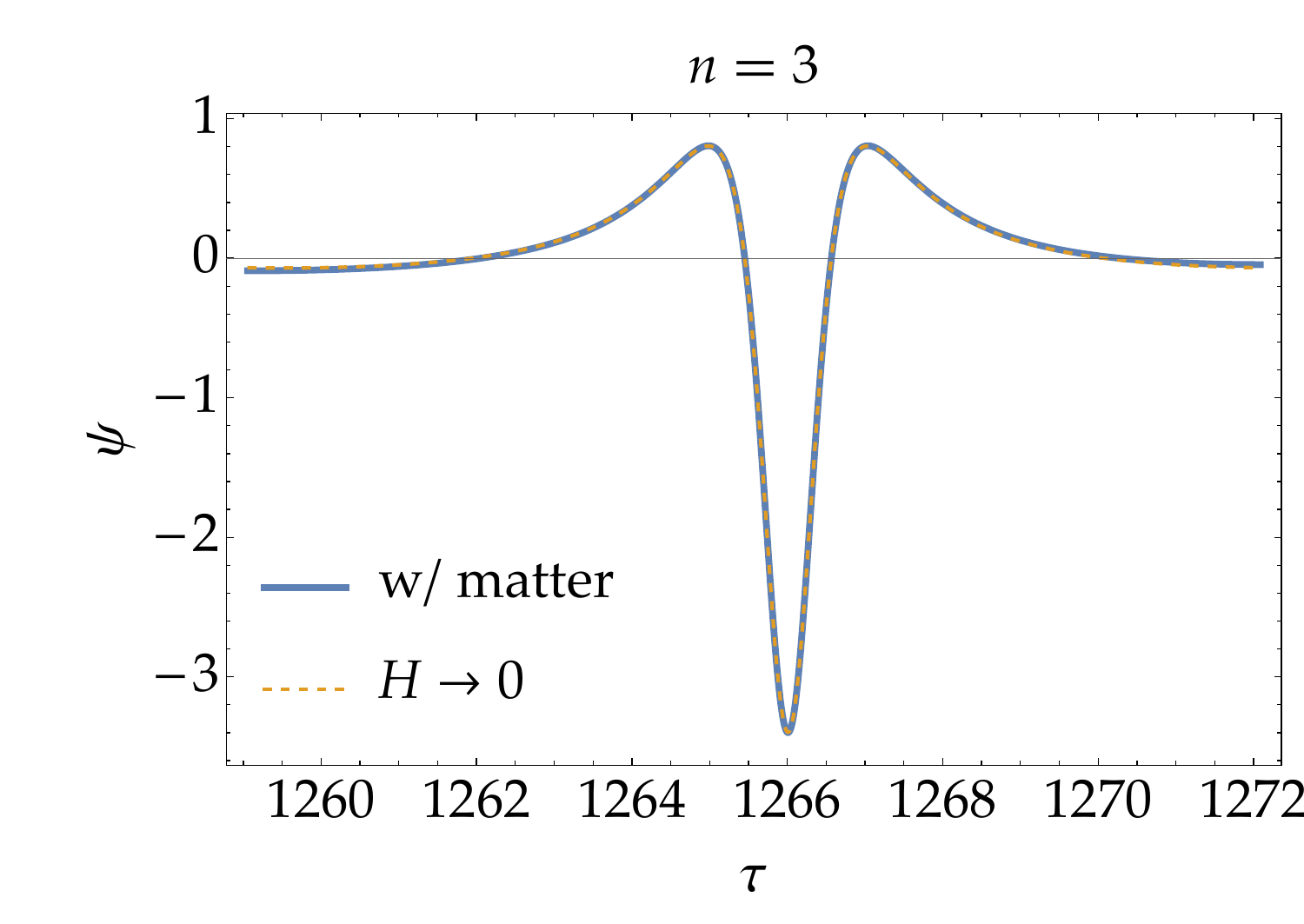}
		\end{minipage}
		\begin{minipage}{0.48\hsize}
			\centering
			\includegraphics[width=\hsize]{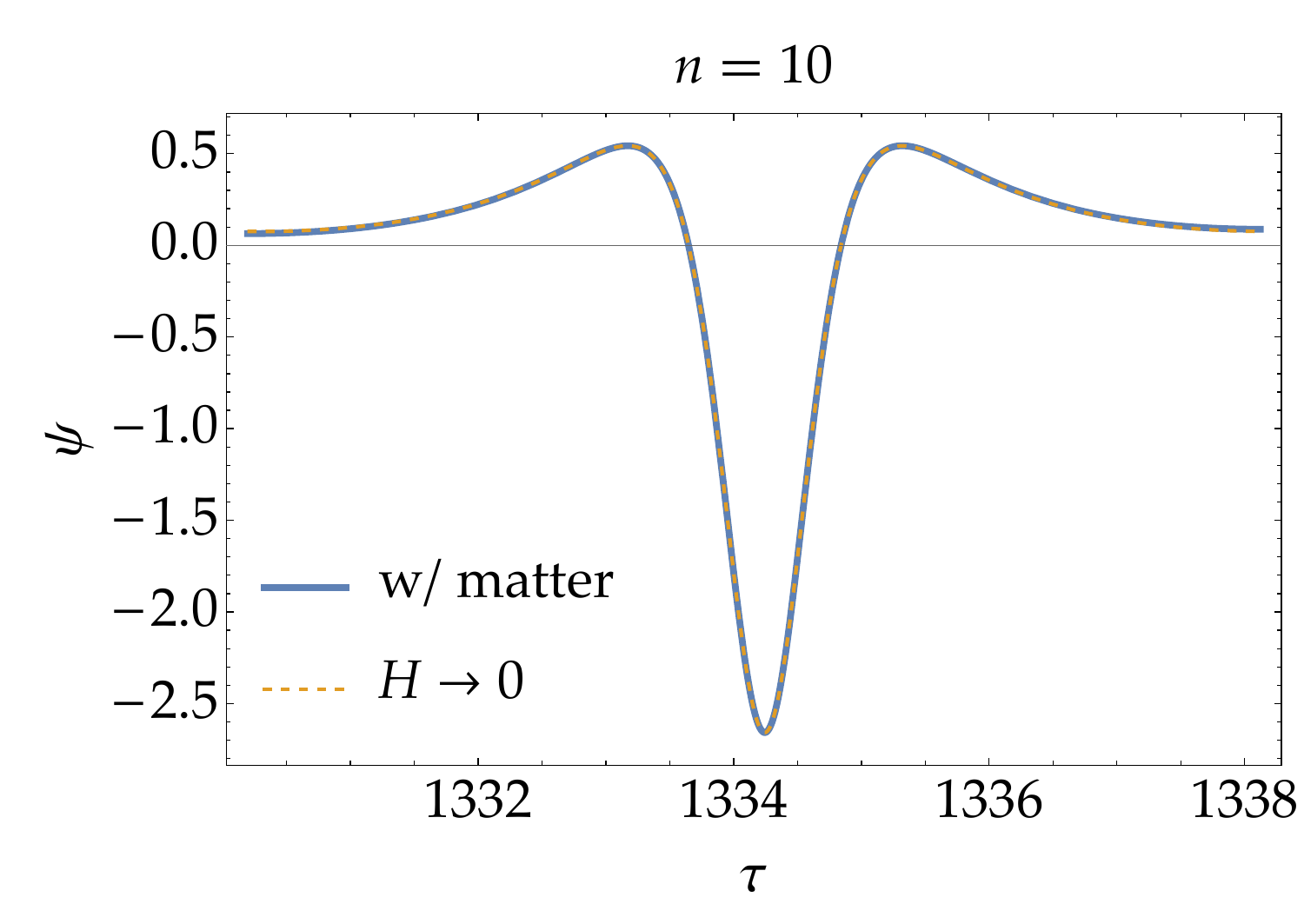}
		\end{minipage}
	\end{tabular}
	\caption{The comparison between the normalised periodic function $\psi_n(\tau)$~\eqref{eq: psin} (blue) in the expanding universe with the cosmological parameters used in the main text, and its counterpart $\psi(\theta(\tau))$~\eqref{eq: psi} (orange dashed) in the $H\to0$ limit with the amplitude $\varphi_\ui$ given by the ``average'' one $\bar{\varphi}_n=\frac{\varphi(\tau_n)+\varphi(\tau_{n+1})}{2}$ for $n=1$ (top left), $2$ (top right), $3$ (bottom left), and $10$ (bottom right).}
	\label{fig: psi with matter}
\end{figure*}

Let $\tau_n$ the $n^\text{th}$ time when $\partial_\tau\varphi(\tau)=0$. The $n^\text{th}$ oscillation segment is defined as the duration from $\tau_n$ to $\tau_{n+1}$.
For this segment, the normalised periodic function (in $\tau$), $\psi_n(\tau)$, can be defined by
\bae{\label{eq: psin}
  \psi_n(\tau)=-\frac{\calV''(\varphi(\tau))-\expval{\calV''}_n}{\sqrt{2}\sigma_n},
}
similarly to Eq.~\eqref{eq: psi}. Here, the time average is defined over the segment:
\bege{
  \expval{\calO}_n=\frac{1}{\tau_{n+1}-\tau_n}\int_{\tau_n}^{\tau_{n+1}}\calO(\tau)\dd{\tau}, \\
  \sigma_n^2=\expval{(\calV'')^2}_n-\expval{\calV''}_n^2.
}
In Fig.~\ref{fig: psi with matter}, we compare $\psi_n(\tau)$ for the cosmological parameters used in the main text with the counterpart $\psi(\theta(\tau))$ in the $H\to0$ limit with the amplitude $\varphi_\amp$ given by $\bar{\varphi}_n\coloneqq\frac{\varphi(\tau_n)+\varphi(\tau_{n+1})}{2}$.
The times through the origin are aligned.
One finds that $\psi_n(\tau)$ is well approximated by $\psi(\theta(\tau))$ for any oscillation segment.
It implies that for each oscillation segment, the cosmic expansion can be neglected and the resonance structure is characterised by the ``average'' amplitude $\bar{\varphi}_n$, and this ``average'' amplitude decreases in time due to the cosmic expansion.

In addition to the redshift of the amplitude, the wavenumber should also be replaced with the physical, redshifting one, $\kappa=(1+z)k/m_\uth$, where $k$ is the comoving wavenumber.
In Fig.~\ref{fig: Floquet chart}, several example redshift flows are overplotted as white arrows.
If they intersect the resonance bands, the corresponding modes get enhanced.

The enhancement of the perturbation can also be checked by directly solving the linearised \ac{EoM} (without metric perturbations)
\bae{
    \delta\ddot{\phi}_k+3H\delta\dot{\phi}_k+\pqty{\frac{k^2}{a^2}+V''(\phi_0(t))}\delta\phi_k=0.
}
With the Bunch--Davies initial condition,
\bae{
  \delta\phi_k\overset{t\to0}{\to}\frac{1}{a\sqrt{2k}} \qc \delta\dot{\phi}_k\overset{t\to0}{\to}-\frac{i}{a^2}\sqrt{\frac{k}{2}},
}
we illustrate the evolution of the power spectrum $\calP_\phi=\frac{k^3}{2\pi^2}\abs{\delta\phi_k}^2$ in  Fig.~\ref{fig: linear power}. 
One can clearly see a significant enhancement. 
It can also be used as a consistency check for the lattice simulations we perform in the main text.

\section{Estimate of initial oscillation period}\label{sec: estimate_period}

In this appendix, we estimate the initial half-period of oscillations of $\phi$. 
To this end, we follow Refs.~\cite{Karam:2021sno, Tomberg:2021bll} (see also Appendix~A of Ref.~\cite{Matsui:2023ezh} for $w(\rho(\phi))$) but generalise their technique taking into account the matter component when possible. 

As in the above references, the analysis is based on two steps: (1) neglect Hubble expansion and consider oscillations within a Hubble time, and then (2) consider the decrease of the energy density by Hubble expansion. Because the matter affects the dark energy dynamics only through the Hubble parameter, it is relevant only to step~(2). Therefore, the results in the literature for step~(1) are intact. 

For step~(1), we refer the reader to the original references for details, but we briefly summarise definitions of key quantities, etc., for convenience.  

All the relevant quantities for the homogeneous dynamics can be derived from the so-called abbreviated action
\begin{align}
    W(\rho) &\equiv \int_{\phi_1}^{\phi_2} \dd{\phi} \eval{\frac{\partial P}{\partial \dot \phi}}_{\dot \phi = \dot \phi (\phi, \rho)} \nonumber \\
    &= 2 \int_0^{\phi_\text{amp}} \dd{\phi} \sqrt{2(\rho - V(\phi))},
\end{align}
where $\phi_1$ and $\phi_2$ are the turning points at which $\dot{\phi} = 0$, and we have assumed canonical normalization and $V(\phi)=V(-\phi)$ in the second equality. The energy density $\rho$ and the pressure $P$ satisfy $\rho = \dot{\phi} \frac{\partial P}{\partial \dot{\phi}} - P$. The energy density $\rho$ is conserved in step (1). 
The half period (in physical time) $\Delta T$ is given by
\begin{align}
\Delta T = \frac{\partial W}{\partial \rho},
\end{align}
and the averaged pressure is given by
\begin{align}
    \bar P = \frac{W}{\Delta T} - \rho.
\end{align}

In step (2), the dark energy density $\rho$ is diluted as 
\begin{align}
    \dot{\rho} + 3 H (\rho + \bar{P}) = 0.
\end{align}
Note that $H$ can be affected by the matter component. Here, $\rho$ and $\bar P$ include only the dark energy component.  
Multiplying $\Delta T$ with this equation, we have $\Delta T \dot{\rho} + 3 H W = 0$. Using the above formula of $\Delta T$, this leads to $\dot{W} + 3 H W = 0$, meaning that $W \propto a^{-3}$. This fact itself is the same as in the literature. 

Within a half period, the dark energy density decreases as 
\begin{align}
    \frac{\Delta \rho}{\rho} \approx \Delta T \frac{\dot{\rho}}{\rho} = - \frac{3 H W}{\rho} = - \frac{W}{\Omega_\phi H}, 
\end{align}
where we have used $\rho = 3 \Omega_\phi H^2$. 
If we set $\Omega_\phi = 1$, it reduces to $- W/H$, reproducing the result in the literature. 
Substituting the concrete expression for $W$ (see Ref.~\cite{Tomberg:2021bll}), it reads
\begin{align}
    \frac{\Delta \rho}{\rho} \approx - 3 \sqrt{2} \pi \frac{H}{\sqrt{V_0}} f .
\end{align} 
When $\Omega_\phi = 1$, it reduces to the expression in the literature. 

We can now discuss the initial amplitude and the corresponding potential value $V(\phi_\text{amp,i})$.
We can write the following relation.
\begin{align}
    \rho = V_0 + \Delta \rho = V(\phi_\text{amp,i}). 
\end{align}
Then, we have
\begin{align}
    \cosh^2 \left( \frac{\phi_\text{amp,i}}{f} \right) \approx \frac{ \sqrt{V_0}}{3 \sqrt{2} \pi H f }.
\end{align}
This is used in Eq.~\eqref{cosh_estimate}. 
Using $\cosh^2 (x) \approx \frac{1}{4} \exp (2 x) $, we obtain
\begin{align}
    \frac{\phi_\text{amp,i}}{f} \approx - \frac{1}{2} \ln \left( \frac{3 \pi  H f}{2\sqrt{2V_0}} \right).
\end{align}
When we set $\Omega_\phi = 1$, it reduces to 
$\frac{\phi_\text{amp,i}}{f}  \approx - \frac{1}{2} \ln (\sqrt{\frac{3}{2}}\frac{\pi}{2} f )$, which is consistent with the literature. 

Next, we evaluate the half period at the initial stage as
\begin{align}
    \Delta T = \frac{\partial W}{\partial \rho} \approx \frac{\pi f}{\sqrt{2 (V_0 - \rho)}}.
\end{align}
This expression is the same as in the case without matter.
Now, we substitute $V_0 - \rho = |\Delta \rho|$.
\begin{align}
    \Delta T \approx \frac{\Omega_\phi^{1/4} \sqrt{\pi f M_P}}{\sqrt{2 \sqrt{6} V_0}},
\end{align}
where we used $V_0 \approx 3\Omega_\phi H^2 M_P^2$. 

Finally, let us comment on the fragmentation time $T_\text{frag}$, when the $\phi$ condensate fragments due to backreaction.  More precisely, the fragmentation time is defined as the time when the energy densities of the homogeneous background and the produced inhomogeneous modes become comparable. It is estimated as~\cite{Tomberg:2021bll}
\begin{align}
    \frac{T_\text{frag}}{\Delta T} \approx \frac{1}{\frac{\phi_\text{amp}}{f} + C} \ln \left( \frac{4.0 \sqrt{V_0}}{(k_\text{peak}/a)^2}  \right) ,
\end{align}
where the subleading log is neglected, and $0.5 \lesssim C \lesssim 1$ according to Ref.~\cite{Tomberg:2021bll}.
If we fix $m_\text{th}$ as in Ref.~\cite{Tomberg:2021bll} (to fit the CMB normalisation in the context of preheating after inflation), we obtain
\begin{align}
    \frac{T_\text{frag}}{\Delta T} \approx \frac{\ln (\Mpl / m_\text{th})}{- \frac{1}{2} \ln (f/\Mpl)} \approx \frac{10}{- \log_{10} f}, \label{T_frag_inflation}
\end{align}
reproducing the result in the reference.
On the other hand, if we fix $V_0$ to explain dark energy, we instead obtain
\begin{align}
    \frac{T_\text{frag}}{\Delta T} \approx \frac{\ln(f/H_0)}{- \frac{1}{2} \ln (f / \Mpl)} \approx \frac{\ln(f/\Mpl) + 138 }{- \frac{1}{2} \ln (f/\Mpl)}. \label{T_frag_DE}
\end{align}
 This estimate is rough, and we observe a discrepancy with an $\mathcal{O}(1)$ factor with our numerical lattice simulation for $f=10^{-3}\Mpl$.  On the other hand, the estimate~\eqref{T_frag_DE} is substantially larger than the above one~\eqref{T_frag_inflation}.  Eq.~\eqref{T_frag_inflation} becoming less than $1$ is one of the reasons that question the extrapolation to smaller and smaller $f$ in Ref.~\cite{Tomberg:2021bll}. This criterion is substantially loosened in our case with Eq.~\eqref{T_frag_DE}. See also the related discussions around Eq.~\eqref{f_min}, which highlight the substantially loosened lower bound on $f$ in our case.

\section{Full MCMC results}
\label{sec: FUll MCMC}

\begin{figure*}
  \centering
  \includegraphics[width=0.95\hsize]{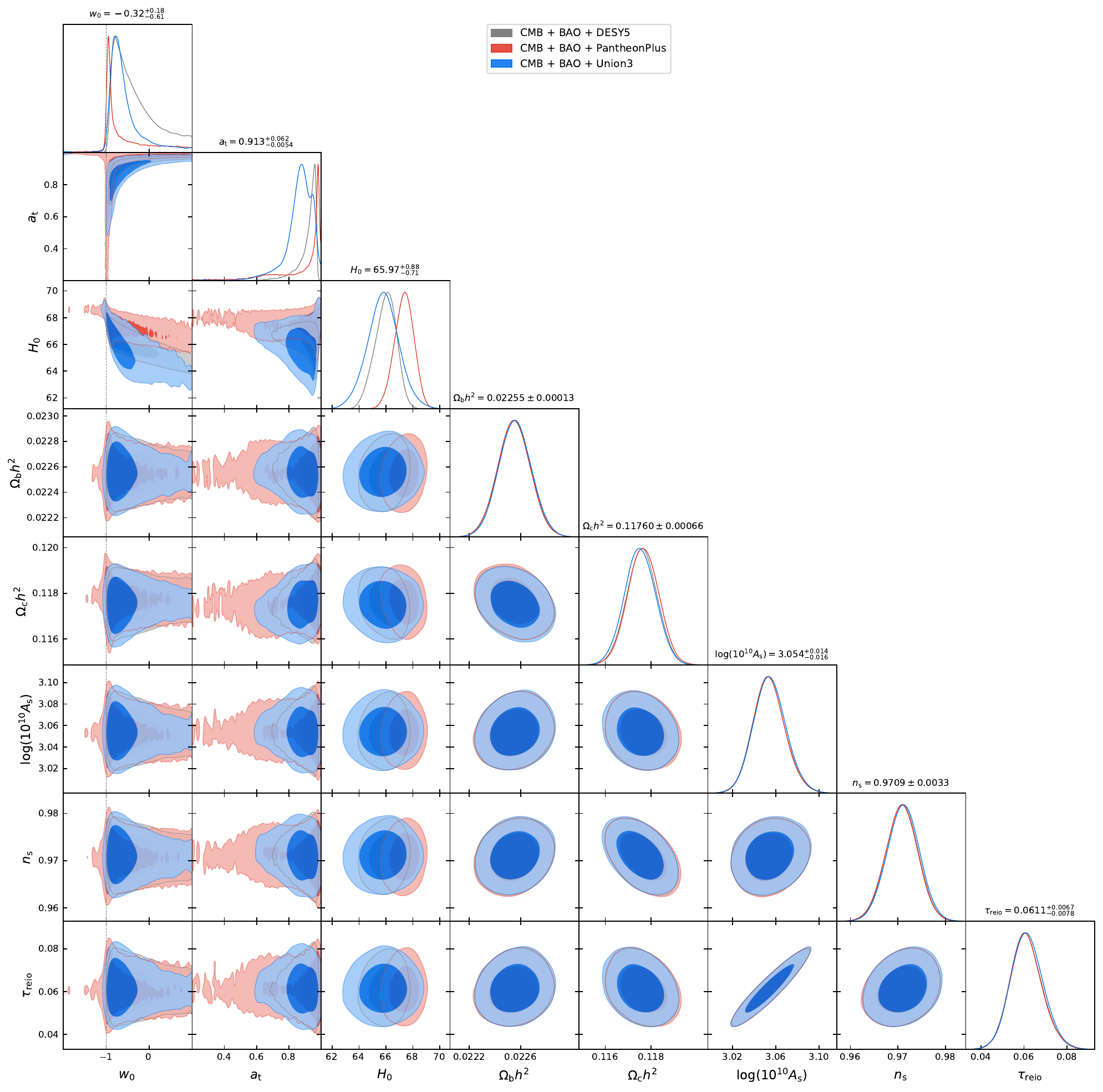}
  \caption{Posterior distribution for the TDE model}
  \label{fig: TDE_MCMC}
\end{figure*}

\begin{figure*}
  \centering
  \includegraphics[width=0.95\hsize]{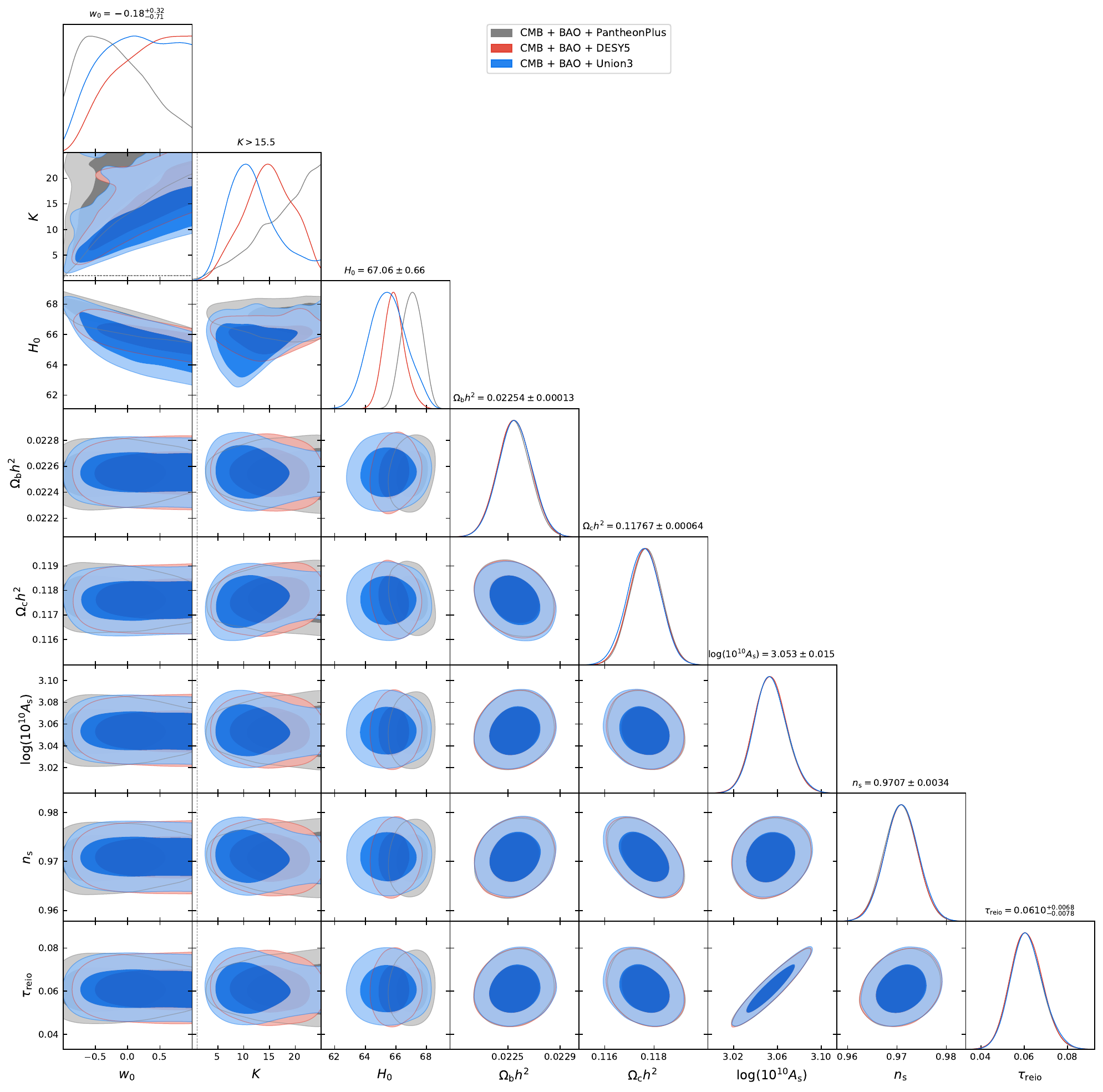}
  \caption{Posterior distribution for the DSCh model}
  \label{fig: DSCh_MCMC}
\end{figure*}

In this appendix, we show the full MCMC results referred to in Sec.~\ref{sec: MCMC analysis}. Flat priors are assumed for all parameters, with the ranges $w_0 \in (-2,1)$ and $a_\mathrm{t} \in (0.2,1)$ for the TDE model and $w_0 \in (-1,1)$ and $K \in (0.01,25)$ for the DSCh model. The comparison of the models is summarised in Table~\ref{tab:full_MCMC_TDE}.  We also plot the 1$\sigma$ posterior distribution for each dark energy model in Fig.~\ref{fig: DESY5_z-wz}. The corner plots of the posterior distributions for the two models are shown in Figs.~\ref{fig: TDE_MCMC} and \ref{fig: DSCh_MCMC}.

\bfe{width=0.95\hsize}{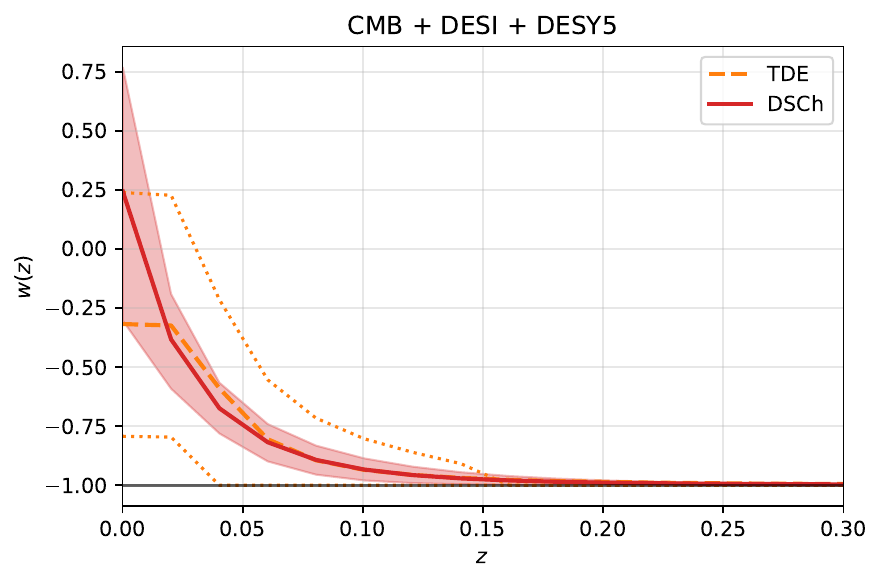}{
1$\sigma$ posterior distributions for the \ac{DSCh} model (red band) and the \ac{TDE} model (region bounded by upper and lower orange-dotted lines), where the red-thick and orange-dashed lines show their best-fit values.
}{fig: DESY5_z-wz}

\begin{table*}
\renewcommand{\arraystretch}{1.3}
\footnotesize
\begin{center}
  \caption{$68\%$ CL posterior for each DE model}
  \label{tab:full_MCMC_TDE}
  \begin{ruledtabular}
  \begin{tabular}{lcccccccc}
    & \quad & \multicolumn{3}{c}{TDE} & \quad & \multicolumn{3}{c}{DSCh}\\
    Parameters & & DESY5 & PantheonPlus & Union3 & & DESY5 & PantheonPlus & Union3\\
    \hline
    $w_0$ & & $ -0.32^{+0.18}_{-0.61}$ & $ -0.54^{+0.21}_{-0.51}$& $ -0.548^{+0.076}_{-0.40}$ & & $ > 0.00415$ &$ -0.18^{+0.32}_{-0.71}$ &$ 0.12^{+0.82}_{-0.34}$\\
    $a_\mathrm{t}$ or $K$ & & $ 0.913^{+0.062}_{-0.0054}$ & $ > 0.853$ & $ 0.861^{+0.10}_{-0.032} $ & & $ 14.8^{+5.5}_{-4.5}$ & $ > 15.5$ & $ 12.0^{+3.1}_{-6.1}$ \\
    $\Omega_\mathrm{b} h^2$ & & $ 0.02255\pm 0.00013$ & $ 0.02255\pm 0.00013$ & $ 0.02256\pm 0.00013$ & & $ 0.02255\pm 0.00013$ & $ 0.02254\pm 0.00013$ & $ 0.02255\pm 0.00013$ \\
    $\Omega_\mathrm{c} h^2$ & & $ 0.11760\pm 0.00066$ & $ 0.11766\pm 0.00068$ & $ 0.11752\pm 0.00068$ & & $ 0.11764\pm 0.00065$ & $ 0.11767\pm 0.00064$ & $ 0.11758\pm 0.00067$ \\
    $100\theta_\mathrm{MC}$ & & $1.04122\pm 0.00028$ & $ 1.04121\pm 0.00028$ & $1.04123\pm 0.00028$ & & $1.04121\pm 0.00028$ & $1.04122\pm 0.00028$ & $1.04122\pm 0.00028$\\
    $\log(10^{10}A_\mathrm s)$ & & $3.054^{+0.014}_{-0.016}$& $3.054^{+0.014}_{-0.015}$ & $3.054\pm 0.015$ & & $3.054\pm 0.015$ & $3.053\pm 0.015$ & $3.054^{+0.014}_{-0.016}$\\
    $n_\mathrm s$ & & $0.9709\pm 0.0033$ & $0.9708\pm 0.0034$ & $0.9711\pm 0.0034$ & & $0.9708\pm 0.0034$ & $0.9707\pm 0.0034$ & $0.9710\pm 0.0034$\\
    $\tau$ & & $0.0611^{+0.0067}_{-0.0078}$ & $0.0611^{+0.0066}_{-0.0078}$ & $0.0615^{+0.0070}_{-0.0079}$ & & $0.0611\pm 0.0074$ & $0.0610^{+0.0068}_{-0.0078}$ & $0.0614^{+0.0067}_{-0.0079}$\\
    \hline
    $H_0/(\mathrm{km/s/Mpc})$ & & $ 65.97^{+0.88}_{-0.71}$ & $ 67.36\pm 0.74$& $ 65.8\pm 1.2$ & & $ 65.89^{+0.59}_{-0.73}$ & $ 67.06\pm 0.66$ &$ 65.5^{+1.1}_{-1.3}$\\
    $ \Delta \chi^2$ & & $ -14.8$& $ -3.66$ & $ -8.68$ & & $ -15.4$ &$ -4.59$ &$ -7.26$ \\
  \end{tabular}
\end{ruledtabular}
\end{center}
\end{table*}

\bibliography{main}    
\end{document}